\documentclass[12pt]{article}

\pdfoutput=1

\usepackage{tikz-cd}
\usetikzlibrary{arrows,snakes,shapes.arrows,decorations.markings}
     \tikzset{>=triangle 90}
     \tikzstyle{bbc}=[draw,circle,fill=black,scale=.75]
     \tikzstyle{rc}=[circle,fill=red,scale=.6]
     \tikzstyle{wc}=[draw,circle,scale=.75]

\usepackage[top=80pt,bottom=85pt,left=85pt,right=85pt]{geometry}
\usepackage{amssymb}
\usepackage{amsmath}
\usepackage{setspace}
\usepackage{sectsty}
\usepackage{cite}
\usepackage{bm}
\usepackage{bbm}
\usepackage[utf8]{inputenc}
\usepackage[T1]{fontenc}
\usepackage{comment}
\usepackage[english]{babel}
\usepackage{afterpage}
\usepackage{bbold}
\usepackage{mathdots}
\usepackage{physics}
\usepackage{makecell,booktabs}
\usepackage{multirow}
\usepackage{lscape}
\usepackage{mathtools}
\usepackage{dashrule}

\usepackage{graphicx,subcaption}
\usepackage{setspace}
\usepackage{float}
\usepackage{booktabs}
\usepackage[debug,pageanchor=false]{hyperref}
\definecolor{link}{rgb}{.8,.15,.1}
\definecolor{pigment}{rgb}{0.36, 0.54, 0.66}
\definecolor{pigment2}{rgb}{0.19, 0.55, 0.91}
\definecolor{pigment3}{rgb}{0.2, 0.2, 0.6}
\definecolor{light-gray}{gray}{0.75}
\hypersetup{colorlinks=true,linkcolor=link,citecolor=link,urlcolor=link,linktocpage}

\usepackage{array,multirow,booktabs,longtable}
\usepackage{mathrsfs}
\usepackage{tikz-cd} 
\usetikzlibrary{backgrounds, arrows,calc,shapes,decorations.pathreplacing, decorations.markings, automata,positioning}
\tikzset{%
  >={Latex[width=2mm,length=2mm]},
            base/.style = {rectangle, rounded corners, draw=black,
                           minimum width=4cm, minimum heigwht=1cm,
                           text centered, font=\sffamily},
  activityStarts/.style = {base, fill=orange!15},
       startstop/.style = {base, fill=orange!15},
    activityRuns/.style = {base, fill=orange!15},
         process/.style = {base, minimum width=2.5cm, fill=orange!15,
                           font=\ttfamily},
}

\newcommand{\red}[1]{}

\tikzset{
        cvertex/.style={circle,draw=black,inner sep=1pt,outer sep=3pt},
        vertex/.style={circle,fill=black,inner sep=1pt,outer sep=3pt},
        star/.style={circle,fill=yellow,inner sep=0.75pt,outer sep=0.75pt},
        tvertex/.style={inner sep=1pt,font=\scriptsize},
        gap/.style={inner sep=0.5pt,fill=white}}

\tikzstyle{mybox} = [draw=black, fill=blue!10, very thick,
    rectangle, rounded corners, inner sep=10pt, inner ysep=20pt]
\tikzstyle{boxtitle} =[fill=blue!50, text=white,rectangle,rounded corners]

\newcolumntype{C}{>{\hfil$}p{3cm}<{$\hfil}}
\newcolumntype{P}{>{\hfil$}p{7.7cm}<{$\hfil}}
\newcolumntype{L}{>{\hfil$}p{2.8cm}<{$\hfil}}
\newcolumntype{S}{>{\hfil$}p{1.8cm}<{$\hfil}}
\newcolumntype{R}{>{\hfil$}p{5.2cm}<{$\hfil}}
\newcolumntype{U}{>{\hfil$}p{4.2cm}<{$\hfil}}
\newcolumntype{Q}{>{\hfil$}p{6.4cm}<{$\hfil}}
\newcolumntype{T}{>{\hfil$}p{1.9cm}<{$\hfil}}
\newcolumntype{V}{>{\hfil$}p{5.8cm}<{$\hfil}}
\newcolumntype{H}{>{\hfil$}p{1.8cm}<{$\hfil}}
\newcolumntype{W}{>{\hfil$}p{3.7cm}<{$\hfil}}
\newcolumntype{A}{>{\hfil$}p{6cm}<{$\hfil}}
\newcolumntype{B}{>{\hfil$}p{2cm}<{$\hfil}}
\newcolumntype{D}{>{\hfil$}p{7.4cm}<{$\hfil}}

\newcommand{\todo}[1]{}
\renewcommand{\todo}[1]{{\color{red} TODO: {#1}}}
\renewcommand{\red}[1]{{\color{red} {#1}}}

\newcommand{\be}{\begin{equation}}  
\newcommand{\ee}{\end{equation}}  
\newcommand{\bea}{\begin{align}}
\newcommand{\eea}{\end{align}}
\newcommand{\bp}{\begin{bmatrix*}[r]}  
\newcommand{\bpp}{\begin{bmatrix}}  
\newcommand{\epp}{\end{bmatrix}}  
\newcommand{\bcd}{\begin{center}
\begin{tikzcd}}
\newcommand{\ecd}{\end{tikzcd} \end{center}}
\newcommand{\bpm}{\begin{pmatrix}}  
\newcommand{\eem}{\end{pmatrix}}

\makeatletter
\@addtoreset{equation}{section}
\makeatother

\begin{document}

\begin{titlepage}

\begin{center}

\vskip .3in \noindent

{\Large \bf{Universal flop of length 1 and 2 \\ \vspace{.5cm} \hspace{1mm} from D2-branes at surface singularities   }}

\bigskip\bigskip\bigskip

Marina Moleti$^{a}$ and Roberto Valandro$^{b}$ \\

\bigskip


\bigskip
{\footnotesize
 \it

$^a$ SISSA and INFN, Via Bonomea 265, I-34136 Trieste, Italy\\
\vspace{.25cm}
$^b$ Dipartimento di Fisica, Universit\`a di Trieste, Strada Costiera 11, I-34151 Trieste, Italy \\
and INFN, Sezione di Trieste, Via Valerio 2, I-34127 Trieste, Italy	
}

\vskip .5cm
{\scriptsize \tt mmoleti at sissa dot it \hspace{1cm} 
    roberto dot valandro at ts dot infn dot it}

\vskip 2cm
     	{\bf Abstract }
\vskip .1in
\end{center}

We study families of deformed ADE surfaces by probing them with a D2-brane in Type IIA string theory.
The geometry of the total space $X$ of such a family can be encoded in a scalar field $\Phi$, which lives in the corresponding ADE algebra and depends on the deformation parameters. 
The superpotential of the probe three dimensional (3d)  theory incorporates a term that depends on the field $\Phi$. By varying the parameters on which $\Phi$ depends, one generates a family of 3d theories whose moduli space always includes a geometric branch, isomorphic to the deformed surface. 
By fibering this geometric branch over the parameter space, the total space $X$ of the family of ADE surfaces is reconstructed. We explore various cases, including when $X$ is the universal flop of length $\ell=1,2$.
The effective theory, obtained after the introduction of $\Phi$, provides valuable insights into the geometric features of $X$, such as the loci in parameter space where the fiber becomes singular and, more notably, the conditions under which this induces a singularity in the total space.
By analyzing the monopole operators in the 3d theory, we determine the charges of certain M2-brane states arising in M-theory compactifications on $X$.

\noindent

\vfill
\eject

\end{titlepage}

\tableofcontents

\newpage

\section{Introduction}

The connection between simple surface singularities and semi-simple Lie algebras, first unveiled through the McKay correspondence in mathematics, is elegantly captured by their shared classification via ADE Dynkin diagrams.
In string theory, this correspondence takes on a richer interpretation. For instance, when compactifying M-theory or Type IIA string theory on a surface with a canonical ADE-type singularity, the resulting lower dimensional theory contains a vector field in the adjoint representation of the corresponding Lie algebra. The Cartan components originate from  the Kaluza-Klein reduction of the three-form $C_3$. The roots of the algebra emerge in a geometric way: the singularity contains a network of two-spheres with vanishing area, whose intersections reflect the structure of the ADE Dynkin diagram. M2-branes or D2-branes wrapping these zero-size spheres produce massless states in the effective theory, charged, via their coupling to $C_3$, under the abelian group generated by the Cartan subalgebra.

This interplay between singularities and Lie algebras can also be studied from the perspective of a probe D2-brane, that is point-like on the singular ADE surface, while extends in three non-compact directions. 
The theories that live on the worldvolume of these D2-branes are described by three-dimensional (3d) $\mathcal{N}=4$ supersymmetric quiver gauge theories, where the quiver diagram has the structure of the corresponding ADE Dynkin diagram. The ADE algebra appears in this theory as a  flavor symmetry. The moduli space  factorizes into a Higgs branch, which has always complex-dimension $2$, and a Coulomb branch, which has complex-dimension $2r$, where $r$ is the rank of the ADE Lie algebra.
The Higgs branch is a copy of the ADE surface; hence, studying these 3d theories is an essential tool  for understanding the geometry of ADE surface singularities and their wide-ranging applications.

Among the many diverse uses of ADE surface singularities, one of the most significant is their fundamental role in the classification of threefold flops in algebraic geometry. Such a classification was given in \cite{Katz:1992aa} by
using the length invariant \cite{Kollar}, which takes on six distinct values $\ell=1,...,6$. 
This invariant has a natural group-theoretical interpretation within the framework of ADE Lie algebras. 
The length classification is accompanied by a construction producing what is known as {\it universal flop of length $\ell$} \cite{Curto:aa}.
This is the  total space of a family of deformed ADE singularities, with types   $A_1, D_4, E_6, E_7, E_8$ and have complex dimension $r+2$, where $r$ is the rank of the associated ADE Lie algebra. 
Any length $\ell$ threefold flop can be constructed via a map to the universal flop of length $\ell$.
In this paper, we focus on $\ell=1,2$. The known examples of threefold flops of $\ell=1$ are the conifold and Reid's pagoda \cite{pagodas}. The most famous $\ell=2$ threefold flops are Laufer \cite{Laufer} and Morrison-Pinkham \cite{pinkham} examples, as well as the most recent Brown-Wemyss threefold \cite{Brown:2017tsa}.

The families of (deformed) ADE surfaces are described as fibrations:  the fiber is a surface where the ADE singularity has been deformed, while the base is given by the space of deformation parameters (collectively named as $\boldsymbol{\varrho}$). At the origin of the parameter space, the surface displays the full ADE singularity, while at a generic point of the base, the surface is smooth. At certain loci in the parameter space, the fiber still develops a singularity governed by a subalgebra of the initial ADE algebra. 
A singularity of the fiber does not imply a singularity in the total space $X_{r+2}$ of the family. However, the latter often harbors singularities of its own, which can sometimes be resolved. In particular, at the origin, the resolution of $X_{r+2}$ might not fully resolve the fiber surface but instead blows up a subset of the spheres that were shrunk at the ADE singularity. This process is known as partial simultaneous resolution.

Recent advances by \cite{Collinucci:2021ofd,Collinucci:2021wty,Collinucci:2022rii,DeMarco:2022dgh} have demonstrated that the geometric data of these families can be encoded in a scalar field $\Phi$ with values in the ADE algebra associated with the surface singularity. The structure of $\Phi$ and its precise dependence on the deformation parameters $\boldsymbol{\varrho}$ dictates how the surface is resolved across the parameter space, including which type of singularity the family develops. This encoding of geometric information into $\Phi$ plays a central role in describing both the local and global features of the ADE-families and in particular of the universal flops of length $\ell$ \cite{Collinucci:2022rii}.

In this paper, we aim to extract the geometric data of the family by analyzing the 3d theory of a D2-brane probing an ADE surface in the presence of a $\Phi$ background. The introduction of $\Phi$ induces a deformation in the superpotential of the D2-brane’s worldvolume theory, that, in many interesting cases,  includes monopole operators. These monopole deformations play a crucial role in understanding the low-energy effective theory but are notoriously challenging to treat. 
The work of \cite{Collinucci:2016hpz,Benini:2017dud,Collinucci:2017bwv}
has provided significant progress in handling these monopole deformations. Their techniques allow for a deeper understanding of the infrared (IR) effective theory that emerges from these monopole deformations, offering insights into the geometric structure of the moduli space. Applying these techniques we were able to write down the quiver and the superpotential for an effective 3d theory, valid at each value of $\boldsymbol{\varrho}$.\footnote{Of course, for a given value of $\boldsymbol{\varrho}$ there might be massive states that can still be integrated out; the relevance of our IR theory is that we have effectively integrated out the fields that are massive for all values of $\boldsymbol{\varrho}$.}
By varying the deformation parameters $\boldsymbol{\varrho}$, the 3d theory and its moduli space change accordingly. This gives rise to a family of three-dimensional theories, each of which features a geometric branch in its moduli space that is isomorphic to the deformed ADE surface.  By fibering this branch over the deformation parameter space $\boldsymbol{\varrho}$, we reconstruct the full family of deformed ADE surfaces. This provides a powerful tool to deal with the families of deformed ADE singularities, particularly in the context of universal flops.

The quiver describing the 3d theories in the family is related to the original quiver, which was dictated by the Dynkin diagram, through operations such as node removal and arrow condensation, which are governed by the action of the monopole deformation. 
If we focus on the maps of the quiver that reproduce the deformed Higgs branches and we set to zero the other maps in a consistent way, we can write down the quiver and the relations that reproduce the ADE family as the moduli space of a given quiver representation. These data match what obtained in \cite{Karmazyn:2017aa}, where the universal flops were reproduced by working out the `contraction algebra' \cite{Donovan_2016}, obtained by operations on the quiver analogous to the one presented in this paper. In particular, for the flop of length $2$, we have been able to reproduce the `universal flopping algebra of length~$2$'  found in \cite{Karmazyn:2017aa}.\footnote{The algorithm that \cite{Karmazyn:2017aa} exposes to obtain the contraction of the quiver and the relations for the ADE families was already introduced in \cite{Cachazo:2001gh} based on more physical grounds.
They allow a specific dependence of the parameter $\boldsymbol{\varrho}$ on an extra field; this dependence enables the integration of the family relations into a 4d superpotential. In this way, \cite{Cachazo:2001gh}  obtains quiver and superpotential for D3-branes probing Calabi-Yau threefold singularities in  Type IIB string theory.}

The 3d gauge theory framework not only recovers the  total space $X_{r+2}$, but also faithfully reproduces its singularities. These singularities arise in specific regions of the parameter space, and the 3d theory provides a natural way to identify and understand these geometric features. 
The appearance of singularities in $X_{r+2}$ is captured in the 3d theory as follows: for generic values of the deformation parameters $\boldsymbol{\varrho}$, the moduli space contains a single branch corresponding to the deformed Higgs branch, inherited from the undeformed theory. However, at special loci in the parameter space, the 3d theory develops additional branches in its moduli space, signaling that the fiber of the family becomes singular at those points. As we said above, these fiber singularities coincide with singularities in the total space $X_{r+2}$ when the simultaneous resolution of $X_{r+2}$ blows up some spheres that are shrunk in the singular surface. In the probe D-brane approach, the sizes of the blown up spheres in the resolution of the family $X_{r+2}$ correspond to the Fayet-Iliopoulos (FI) parameters in the effective 3d theory. At the point $\boldsymbol{\varrho}$ where new branches emerge, the $U(1)$ gauge factor responsible for the FI-term may or may not be broken: if it is preserved, the simultaneous resolution occurs, and $X_{r+2}$ becomes singular at that point.

Finally, we leverage the ADE families to construct spaces where M-theory/Type IIA string theory can be compactified. In particular, taking the universal flop $X_6$ of $\ell=2$ and letting $\boldsymbol{\varrho}$ be functions of a single complex coordinate, one obtains the Morrison-Park threefold \cite{Morrison:2012ei}. Compactifying M-theory/Type IIA on this threefold yields M2/D2-states with both charge $1$ and $2$ with respect to a given $U(1)$ group. This interesting fact was extensively utilized in the F-theory community. 
At the end of this paper, we show how the knowledge of the monopole operators of the effective 3d theory associated with $X_6$ allows to gain information about the charges of M2/D2-states in M-theory/Type IIA compactifications on certain Calabi-Yau (CY) threefolds constructed from $X_6$.

In this work, we provide a detailed analysis of the effective 3d theories that describe the universal flops of length 1 and 2, successfully characterizing their well-known geometric structures. The more intricate cases with $\ell > 2$ are left for future investigation.
The paper is organized as follows. In Section~\ref{Sec:ADEfamilies} we review the ADE families and their description via the field $\Phi$. Section~\ref{Sec:D2ADEsing}  summarizes the key aspects of the 3d theories on D2-branes probing ADE singularities that are relevant to our analysis. In Section~\ref{Sec:D2deformedADEfamilies} we describe the effect on these theories of turning on a background for $\Phi$ and we construct the universal flop of length $1$ from a family of deformed $A_1$ singularities with simultaneous resolution; we also study the effective theory connected to a specific $A_3$ family in preparation for the more complicated $\ell=2$ case. Section~\ref{Sec:UnFlL2} presents our main result: we derive the universal flop of length $2$ and its geometric features, from the family of 3d theories obtained through monopole deformation of a world-volume theory on a D2-brane probing a $D_4$ singularity. In Section~\ref{Sec:Conclusions} we provide our conclusions.

\section{ADE families from the adjoint complex scalar $\Phi$}\label{Sec:ADEfamilies}

\subsection{Families of ADE surfaces}

Consider an ALE surface $S_{\rm sing}$ given by an equation \(f(x,y,z)=0\)  in \(\mathbb{C}^3\) and that has one of the following ADE singularities: 
\begin{equation}\label{ADE singularities}
\begin{split}
& A_n: \hspace{1cm} x^2 + y^2 + z^{n+1} = 0\:, \\
& D_n: \hspace{1cm} x^2 + z y^2 + z^{n-1} = 0\:, \\
& E_6: \hspace{1cm} x^2 + y^3 + z^4 = 0\:, \\
& E_7: \hspace{1cm} x^2 + y^3 + yz^3 = 0\:, \\
& E_8: \hspace{1cm} x^2 + y^3 + z^5 = 0\:. \\
\end{split}
\end{equation}

The surface $S_{\rm sing}$ can be desingularized to a smooth surface $S_{\rm def}$ by deforming the equation as
\begin{equation}\label{Eq:deformedADE}
f (x,y,z)+ \sum_{i=1}^r \mu_i \,g_i(x,y,z) = 0,
\end{equation}
where $r$ is the rank of the ADE algebra, $\mu_i$ are complex parameters and \(g_i\) are monomials belonging to the ring\footnote{When, the monomials belong to the Jacobian ideal \eqref{eq:JacIdeal}, the deformation \eqref{Eq:deformedADE} is called a {\it versal} deformation of the singularity.}
\begin{equation}\label{eq:JacIdeal}
R = \frac{\mathbb{C}(x,y,z)}{\left(\frac{\partial f}{\partial x}, \frac{\partial f}{\partial y}, \frac{\partial f}{\partial z}\right)} \:.
\end{equation}
This deformation blows up $r$ (non-holomorphic) 2-spheres that intersect with a pattern given by the Dynkin diagram pertaining to the corresponding ADE algebra, where the spheres are identified with the {\it simple roots} of the Lie algebra.\footnote{With an abuse of terminology, we will refer to the shrinking 2-spheres as `simple roots'.}
The ADE singularity can also be resolved by a birational map that substitutes the singular point with a collection of complex curves isomorphic to $\mathbb{CP}^1$ (topologically  2-spheres), again intersecting with the pattern given by the Dynkin diagram.\footnote{Notice that the smooth ALE surface is a hyperK\"ahler space and by a rotation of the complex structures one can map a deformation to a resolution. This is the reason  why resolution and deformation give the same differentiable manifold. The algebraic description in \eqref{ADE singularities} of such surfaces fixes the complex structure and then make a clear distinction on what we mean as a resolution or a deformation.}
In both cases, the singular limit corresponds to sending the size of the 2-spheres to zero. 

Let us  focus on the deformation of the ADE singularities.
By treating the \(\mu_i\)'s  as coordinates, the equation \eqref{Eq:deformedADE} defines a space $X_{r+2}$ as an hypersurface in $\mathbb{C}^{r+3}$. It has the structure of a fibration, where the fiber is the (deformed) ADE surface $S_{\rm def}$ and the base is the space spanned by $\mu_i$.  At the origin $\mu_i=0$, the fiber has the full ADE singularity, i.e. all the $r$ simple roots have zero size. 
The base space can be identified with the space \(\mathfrak{t}/\mathcal{W}\), 
where \(\mathfrak{t}\) is the Cartan torus of the Lie algebra $\mathfrak{g}$ and \(\mathcal{W}\) is the Weyl group (see \cite{Katz:1992aa}). Its coordinates \(\mu_i\)'s can then be expressed as \(\mathcal{W}\)-invariant functions of the coordinates~$t_i$ on~\(\mathfrak{t}\).\footnote{For \(\mathfrak{g}=A_{n-1}\), \(\mu_i\) is the \(i\)-th elementary symmetric polynomial in the eigenvalues of an element of the Lie algebra.}

The space $X_{r+2}$ defined by the equation \eqref{Eq:deformedADE} is smooth. However, by performing a \emph{base change}, one can create a singular space whose resolution blows up a subset of the roots of the central ALE fiber. 


To explain the setup, let us consider an ALE family of type \(A_{1}\)  \cite{Katz:1992aa}. The defining equation for the \(A_{1}\) singularity is\footnote{Derived from \eqref{ADE singularities} by a change of variables.}
\be
uv=z^2\:.
\ee
The equation for the family $X_3$ is
\be 
 uv = z^2 + \mu  \qquad\subset\qquad \mathbb{C}^4_{u,v,z,\mu}\:,
\ee
that is a smooth three-dimensional space.
If we make the base change $\mu=-t^2$, i.e. we pull-back the ALE fibration w.r.t. the map
\begin{equation}\begin{split}
\mathfrak{t} &\longrightarrow \mathfrak{t}/\mathbb{Z}_2\\
t & \mapsto \mu=-t^2\,,
\end{split}\end{equation}
where  $\mathfrak{t} \cong \mathbb{C}$ for $\mathfrak{g}=A_1$, we obtain a three-dimensional space with a conifold singularity:
\begin{equation}\label{conifold}
uv = z^2 - t^2 \:.
\end{equation} 
The origin of the family is at $t=0$, where the surface develops an $A_1$ singularity $uv=z^2$. At $t\neq 0$ the singularity of the surface is deformed.
The (small) resolution of the conifold blows up the simple root of \(A_1\) in the central fiber; this is known as a \emph{simultaneous resolution}. The family is now fibered over the base \(\mathfrak{t}\).

Consider a generic ADE algebra of rank $r$. Analogously to the $A_1$ case, we can make a base change by pulling back the ADE family with respect to the map  (see \cite{Katz:1992aa})
\begin{equation}
\begin{aligned}
\mathfrak{t} &\longrightarrow B_\mu\equiv  \mathfrak{t}/\mathcal{W} \\ t_i & \mapsto \mu_j=\mu_j(t)
\end{aligned}
\end{equation}
by taking $\mu_j$'s  as $\mathcal{W}$-invariant functions of coordinates $t_i$ on  $\mathfrak{t}$. This resolves all the simple roots of the ADE algebra in the central fiber.
One can also make different choices for the base change where
\begin{equation}\begin{aligned}
\mathfrak{t} &\longrightarrow B_\varrho \equiv \mathfrak{t}/\mathcal{W}' \\ 
t_i & \mapsto \varrho_k=\varrho_k(t)  \\
\end{aligned}\end{equation}
with $\mathcal{W}'\subset \mathcal{W}$. This subset of the Weyl group is determined by a choice of simple roots, call them
\begin{equation}\label{eq:blownUpRoots}
 \alpha_1,...,\alpha_\kappa\qquad \mbox{with} \qquad \kappa\leq r\:;
\end{equation}
$\mathcal{W}'$ is generated by the reflections under $\alpha_{\kappa+1},...,\alpha_r$. One can visualize it by coloring the roots in \eqref{eq:blownUpRoots}.
The resolution of the family now blows up only these simple roots in the central fiber. This is called a {\it partial simultaneous resolution}. The base of the fibration, that we call $B_\varrho$, is now parametrized by the $r$ $\mathcal{W}'$ invariants~$\varrho_i$~($i=1,...,r$). 
To summarize, the total space of an ADE-family $X_{r+2}$ has the fibration structure
\begin{equation}
   \begin{array}{ccc}
    S_{\rm def} & \hookrightarrow & X_{r+2} \\ && \downarrow \\ && B_\varrho \\
   \end{array}
\end{equation}
where $S_{\rm def}$ is the deformed ADE surface.

\subsection{Universal flops}

The families of ADE surfaces with (partial) simultaneous resolution are the starting point for constructing  three-dimensional CY spaces with simple flops. 
These spaces are sometimes called {\it simple threefold flops}. They have a point-like singularity, whose resolution blows up a single  $\mathbb{CP}^1$. The conifold is the most famous example.  

These threefolds can be mapped into families of ADE surfaces whose simultaneous resolution blows up only one of the simple roots of the corresponding ADE singularity~\cite{Katz:1992aa}. This map is such that the exceptional curve of the threefold is mapped to the blown up root in the ADE family.
The ADE families that blow up a single $\mathbb{CP}^1$ are known as {\it universal flops} \cite{Curto:aa}; they will be the main subject of this paper.

The easiest example comes from the $A_1$ family with simultaneous resolution in equation \eqref{conifold}. This  blows up a single $\mathbb{CP}^1$ and is already a threefold (the map is trivial, $t=w$, giving the equation $uv=z^2-w^2$): it is the well known conifold, that is the easiest example of simple flop threefold.
The next example is a different map that is $t=w^k$ obtaining  the following threefold: $uv=z^2-w^{2k}$. This is known as the Reid's Pagoda threefold. It is distinguished from the conifold; in particular it has a different normal bundle ($\mathcal{O}\oplus \mathcal{O}(-2)$ instead of $\mathcal{O}(-1)\oplus\mathcal{O}(-1)$).

This way of constructing threefold flops gives a simple interpretation of the length invariant $\ell$ that characterizes the threefold flops \cite{Kollar}: it corresponds to the dual Coxeter label of the node of the Dynkin diagram that is being resolved by the partial simultaneous resolution. It is then clear that $\ell$ can take all values from $1$ to $6$ (as it was first proved in \cite{Katz:1992aa}). The previous two examples have indeed length $1$. To provide examples of higher length one needs to consider families of surfaces of type $D$ or $E$ (their normal bundle is $\mathcal{O}(1)\oplus\mathcal{O}(-3)$). 

\begin{figure}[t]
  \begin{center}
     \includegraphics[scale=0.15]{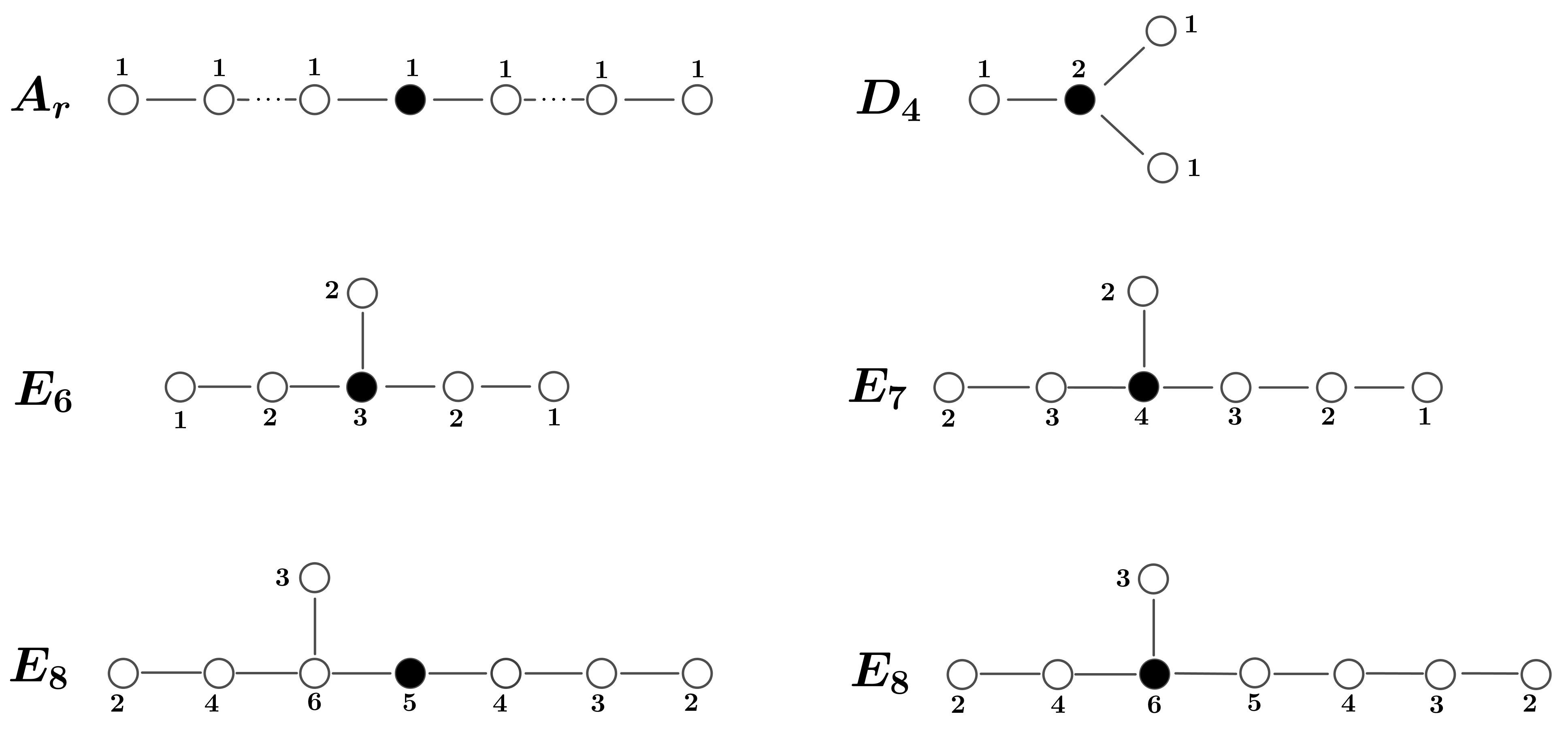}
  \end{center}
  \caption{ADE Dynkin diagrams and dual Coxeter labels of the nodes. The colored node corresponds to the root that is blown up in the (partial) simultaneous resolution in the universal flops. The label of the colored node is the length associated with the corresponding flop.}\label{FigADESimpleFlops}
  \end{figure}

In this paper, we will concentrate on the universal flop of length one and two. They are realized as families of respectively $A_1$ and $D_4$ surfaces. The first one is the $A_1$ family with simultaneous resolution that we have just discussed. The second one is a $D_4$ family, whose simultaneous resolution blows up only the central node of the $D_4$ Dynkin diagram. See Figure~\ref{FigADESimpleFlops}, where we show the blown up root also for the flops of higher length.

\subsection{ADE families from the Higgs field $\Phi$}\label{sec:FamilyFromHiggs}

In \cite{Collinucci:2021ofd,DeMarco:2021try,Collinucci:2022rii,DeMarco:2022dgh} a new way to characterize the ADE families and the simple flops was introduced. It is based on a physical approach to these manifolds, that relies on considering them as spaces for string theory compactifications. Here we briefly review this method.

Type IIA string theory on $R^{1,5}\times S_{\mathfrak{g}}$ engineers a 6-dimensional gauge theory with ADE algebra $\mathfrak{g}$. When the algebra $\mathfrak{g}$ is of type A or D, 
the string setup is dual to type IIA on $R^{1,5}\times S^1\times \mathbb{C}\times\mathbb{R}$ with a stack of D6-branes wrapping the $\mathbb{R}^{1,5}\times S^1$ factor and  with gauge group of type $A$ or $D$ (in the second case, one also has an orientifold O6-plane). 

The duality, known also as 9-11 flip, works in an easy way.  Type IIA on $R^{1,5}\times S_{\mathfrak{g}}$ is trivially uplifted to M-theory on $R^{1,5}\times S^1\times S_{\mathfrak{g}}$. 
The surface $S_{\mathfrak{g}}$ has a natural $\mathbb{C}^*$-fibration over $\mathbb{C}$; it can be used to reduce the M-theory background to type IIA now on $R^{1,5}\times S^1 \times \mathbb{C} \times \mathbb{R}$,  with D6-branes located on points of $ \mathbb{C} \times \mathbb{R}$ where the $\mathbb{C}^*$ fiber degenerates. 

On both sides of the duality the string background engineers a 6d supersymmetric gauge theory with gauge algebra $\mathfrak{g}$. This theory includes an adjoint complex scalar field $\Phi$. From the point of view of type IIA with D6-branes, the vacuum expectation value (vev) of $\Phi$ controls the deformation of the D6-branes along the $\mathbb{C}$ transverse to their worldvolume. By 9-11 flip these are mapped to geometric deformations of the ADE surface. Hence the parameters $\mu_i$ should be related to gauge invariant functions of $\Phi$. 
One can generalize the relation between the gauge theoretic data $\Phi$ and the deformation of the surface $S_\mathfrak{g}$ to the $E$ cases, where a perturbative D-brane interpretation is lacking~\cite{Collinucci:2022rii}.

In the $A_n$ case, giving a vev to $\Phi$ deforms the  locus where the $n+1$ D6-branes live from $z^{n+1}=0$ to
\begin{equation}
 \mbox{D6-brane locus:}\qquad \det\left(z\, \mathbb{1}_{n+1} - \Phi\right)=0\:.
\end{equation}
The deformed surface equation is then
\begin{equation}\label{Eq:DefSAn}
 uv= \det\left(z\, \mathbb{1}_{n+1} - \Phi\right)\:,
\end{equation}
where the $\mathbb{C}^*$ fibration is manifest and the fiber degenerates at the location of the D6-branes.

For the $D_n$ case one has to carefully take into account the presence of the orientifold projection, and the resulting deformed surface equation is \cite{Collinucci:2021wty,Collinucci:2021ofd,Collinucci:2022rii}:
\begin{equation}\label{Eq:DefSDn}
x^2+zy^2-\frac{\sqrt{\text{det}(z\mathbb{1}_{2n}+\Phi^2)}-\text{Pfaff}^2(\Phi)}{z}+2y\hspace{0.1cm}\text{Pfaff}(\Phi)=0\:.
\end{equation}

\subsubsection*{$A_1$-family and the Higgs field $\Phi$}

Let us consider again the $A_1$ case, where there is only one deformation parameter. One can reproduce the deformed equation by two choices~of~$\Phi$, 
\begin{equation}\label{eq:PhiA12choices}
\Phi=\begin{pmatrix}
 t &0\\0&-t\\
\end{pmatrix} \qquad \mbox{or} \qquad
\Phi=\begin{pmatrix}
 0 &1\\\mu&0\\
\end{pmatrix}\:,
\end{equation}
that, through \eqref{Eq:DefSAn}, give the two following equations:
\begin{equation}
uv=z^2-t^2 \qquad\mbox{or} \qquad uv=z^2+\mu\:.
\end{equation}
We recognize that the first one is the singular family with simultaneous resolution, while the second one is the smooth one. For both cases, at the origin of the base, the fiber is the singular $A_1$ surface.

Let us interpret it in type IIA on flat space with D6-branes. The stack of D6-branes supports an $SU(2)$ gauge group when $\Phi=0$. For generic values of the deformation parameter, $\Phi$ as in \eqref{eq:PhiA12choices} breaks this group respectively to $U(1)$ and to nothing. At the origin: 
1) in the first case one recovers the $SU(2)$ gauge group as $\Phi(t=0)=0$; 2)~in~the second case $\Phi$ 
is an upper triangular matrix that breaks $SU(2)$ completely; this second background was named {\it T-brane} in \cite{Cecotti:2010bp}.  Hence, in the second case, for all values of the deforming parameter, the $SU(2)$ is broken, it is never restored. 

In the dual picture, in both cases at the origin of the parameter space we have the same  singular geometry; in fact on this side of the duality (that is analogous to M-theory on the singular $A_1$ surface) the T-brane background is not detectable by the geometric data and one should supplement the background with the T-brane data. In addition to the several previous proposals \cite{Anderson:2013rka,Collinucci:2014qfa,Collinucci:2014taa}, in \cite{Collinucci:2016hpz,Collinucci:2017bwv} the authors defined the T-brane as a specific effect on the worldvolume theory of a probe D2-brane. 

The T-brane background is known to obstruct some resolution of the ADE surface~\cite{Anderson:2013rka,Collinucci:2014taa,Collinucci:2016hpz} . This is compatible with what happens in the second case: 
at the origin ($\mu=0$) there is still a T-brane background $\Phi\neq 0$; this obstructs the resolution of the simple root, that means there is no simultaneous resolution of that root in the family.
This is how the choice of field $\Phi$ at the origin selects one partial simultaneous resolution over another.

\subsubsection*{Generic ADE-families, $\Phi$ and the partial simultaneous resolutions}

Analogous to the $A_1$ case studied above, the various possible (partial) simultaneous resolutions of a family of ADE surfaces can be encoded in the different forms of the field~$\Phi$. This was first described in \cite{Collinucci:2022rii}. Given a choice of partial resolution, the corresponding field $\Phi$ is determined in the following way: 
\begin{itemize}
\item Choose the subset of simple roots $\alpha_1,...\alpha_\kappa$ ($\kappa\leq r$) that are blown up at the origin.\footnote{In the $A_1$ case the only options are either to resolve the single  simple root or resolving none. In the $A_2$ case, for example, one has the following three choices: 1) resolve both the simple roots, 2) resolve one of the simple roots and 3) resolve none.} 
\item Take the abelian subalgebra of $\mathfrak{g}$ generated by
\begin{equation}\label{calH}
\mathcal{H}=\langle\alpha_1^*,...\alpha_\ell^*\rangle\:,
\end{equation}
where  $\alpha_1^*,...,\alpha_r^*$ is a basis of the Cartan subalgebra dual to the set of simple roots.
\item The commutant of $\mathcal{H}$ in $\mathfrak{g}$ is a Levi subalgebra $\mathcal{L}$ of $\mathfrak{g}$ that takes the form
\begin{equation}\label{PhiInL}
\mathcal{L}= \bigoplus_{h} \mathcal{L}_h \oplus \mathcal{H} \:,
\end{equation}
with $\mathcal{L}_h$ simple Lie algebras.
\item $\Phi$ must be a generic element of $\mathcal{L}$.\footnote{
This can be understood in the  type IIA picture with D6-branes \cite{Collinucci:2021ofd}: the resolution corresponds here to the motion of the D6-branes along the direction $\mathbb{R}$ in $\mathbb{R}^{1,5}\times S^1\times \mathbb{C}\times \mathbb{R} $ and are then a Cartan vev for a real adjoint field $\phi\in \mathcal{H}$. The equations of motion impose $\phi$ to commute with $\Phi$. Hence $\Phi$ must live in the commutant of $\mathcal{H}$.
}  
\item The Casimir invariants of $\Phi$ tell us how the ALE fiber is deformed. Since at the origin of $\mathfrak{t}/\mathcal{W}'$ the fiber presents the full ADE singularity, all the Casimir invariants of $\Phi$ must vanish when $\varrho_i=0$ ($i=1,...,r$), i.e. $\Phi(\varrho=0)$ should be a  nilpotent element of $\mathcal{L}$. 
In particular, when restricted on each summand $\mathcal{L}_h$ of the Levi subalgebra, $\Phi$ must be in the corresponding principal nilpotent orbit \cite{Collinucci:2022rii}.\footnote{This will prevent the family to develop singularities that cannot be resolved. In the $A$ and $D$ cases this requirement translates to asking that $\Phi(\varrho)$ is a reconstructible Higgs \cite{Cecotti:2010bp}, i.e. it is written in terms of the Casimir invariants of the $\mathcal{L}_h$'s.} 

\item The Higgs field $\Phi(\varrho)$ must deform the singularity outside the origin. The coordinates $\boldsymbol{\varrho}$ should parametrize a  transverse direction to the nilpotent orbit (that includes $\Phi(0)$) in $\mathfrak{g}$. This is done by taking $\Phi(\varrho)$ in the Slodowy slice\footnote{See for example Appendix B of \cite{DeMarco:2022dgh} for a definition of the Slodowy slice.} in $\mathfrak{g}$ that passes through $\Phi(0)$. This allows to give a canonical form for the Higgs field~$\Phi(\varrho)$. 

Let us consider  $\mathcal{L}_h=A_{m-1}$ as an example that will be useful in the following. The $\Phi$ built as above is then in the form of a \emph{reconstructible Higgs}  \cite{Cecotti:2010bp}:
\be\label{recHiggsU}
\Phi|_{A_{ m-1}} =\left(\begin{array}{cccccc}
0 & 1 & 0 & \cdots && 0 \\
\varrho_1 & 0 & 1 & \ddots && \vdots\\
\vdots & \ddots & \ddots & \ddots && 0  \\
 \varrho_{m-2} & \cdots  & \varrho_1 & 0 && 1 \\
\varrho_{m-1} & \varrho_{m-2} & \cdots & \varrho_1 && 0  \\
\end{array} \right)\:,
\ee
with $\varrho_j$ ($j=1,...,m-1$) that are Casimir invariants of $\Phi|_{A_{ m-1}} $ (we called them the partial Casimir of $\mathfrak{g}$). 
There are analogous canonical forms when the summand is a different Lie algebra. 
Collecting the Casimirs  $\varrho_j$'s for each summand $\mathcal{L}_h$ and the coefficient deformations along $\mathcal{H}$ one obtains the set of invariant coordinates  $\boldsymbol{\varrho}$ that span the base of the family with simultaneous partial resolution.
The total Casimirs of $\Phi$ that appear as coefficients of the deforming monomials in the versal deformation of the ADE singularity, will be functions of $\boldsymbol{\varrho}$.

\end{itemize}

In the previous $A_1$ example, the two possible choices for the Levi algebra are
\begin{equation}
\mathcal{L} = \langle \alpha^* \rangle \qquad \mbox{ and } \qquad \mathcal{L} = \mathfrak{g}_{A_1} \,,
\end{equation} 
i.e. in the first case $\mathcal{L}$ is the Cartan subalgebra, while in the second case, it is the whole~$A_1$ algebra. In the second case, one has a simple summand and $\Phi$ takes the form of a reconstructible Higgs as can be seen in \eqref{eq:PhiA12choices}.

\section{D2-branes probing ADE singularities}\label{Sec:D2ADEsing}

The starting point of our analysis is the theory leaving on the worldvolume of a D2-brane probing an ADE singularity. In this section, we will review its relevant features.

\subsection{The three-dimensional theory and its moduli space}\label{sec:susy3dtheories}

The worldvolume theory of a D2-brane probing and ADE singularity (see \eqref{ADE singularities}) is a $\mathcal{N}=4$ 3d supersymmetric quiver gauge theory. 

\begin{figure}[H]
 \centering
\begin{tikzpicture} [place/.style={circle,draw=black!500,fill=white!20,thick,
inner sep=0pt,minimum size=10mm}, transition/.style={rectangle,draw=black!50,fill=black!20,thick,
inner sep=0pt,minimum size=10mm}, decoration={ markings,
mark=between positions 0.35 and 0.85 step 2mm with {\arrow{stealth}}}]
 \node at ( -5,0) [place] (1) {$1$};\node at (-5, -0.8) {$\phi_2$};\node at (-4, -0.7) {$q_{2}$};\node at (-4, 0.35) {$\tilde{q}_{2}$};
 \node at ( 5,0) [place] (4) {$1$}; \node at (5, -0.8) {$\phi_{r+1}$};\node at (4, -0.7) {$q_{r}$};\node at (4, 0.35) {$\tilde{q}_{r}$};
  \node at ( 3,0) [place] (3) {$1$}; \node at (3, -0.8) {$\phi_r$};\node at (3, 3.4) {$q_{r+1}$};\node at (3, 2.2) {$\tilde{q}_{r+1}$};
  \node at (- 3,0) [place] (2) {$1$}; \node at (-3, -0.8) {$\phi_3$};
  \node at (0,4)[place] (0) {$1$}; \node at (0, 4.8) {$\phi_1$};\node at (-3, 3.3) {$q_{1}$};\node at (-3, 2) {$\tilde{q}_{1}$};
   \node at (-1.5,0) (5) {$$}; \node at (-2, -0.7) {$q_{3}$};\node at (-2, 0.35) {$\tilde{q}_{3}$};
    \node at (1.5,0) (6) {$$}; \node at (2, -0.7) {$q_{r-1}$};\node at (2, 0.35) {$\tilde{q}_{r-1}$};
     \node at (0,0) (7) {$\cdots$}; 
 \draw[->, -Stealth, thick] (0)[bend right] to (1);  \draw[->,-Stealth,thick] (1)--(0);\draw[->,-Stealth,thick] (4)[bend right] to (0);\draw[->,-Stealth,thick] (0)--(4);
 \draw[thick, -Stealth] (4)--(3);\draw[thick, -Stealth] (3)[bend right] to(4);
 
 \draw[thick,Stealth-] (1)--(2);\draw[thick, Stealth-] (2)[bend left] to(1);
 
 \draw[thick, Stealth-] (2)--(5);\draw[thick, Stealth-] (-1.5, -0.2)[bend left] to(2);
 
 \draw[thick, -Stealth] (3)--(6);\draw[thick, -Stealth] (1.5, -0.2)[bend right] to(3);
\end{tikzpicture}
\caption{$A_r$ theory. For each node $i$, $i=1,...,r+1$, there is a $\mathcal{N}=4$ $U(1)$ vector multiplet $V_i$ containing a $\mathcal{N}=2$ vector multiplet and an adjoint chiral $\phi_i$. Pairs of oriented lines between adjacent nodes represent bifundamental hypermultiplets $(q_i, \tilde{q}_i)$.}\label{Fig:ArQuiver}
\end{figure}
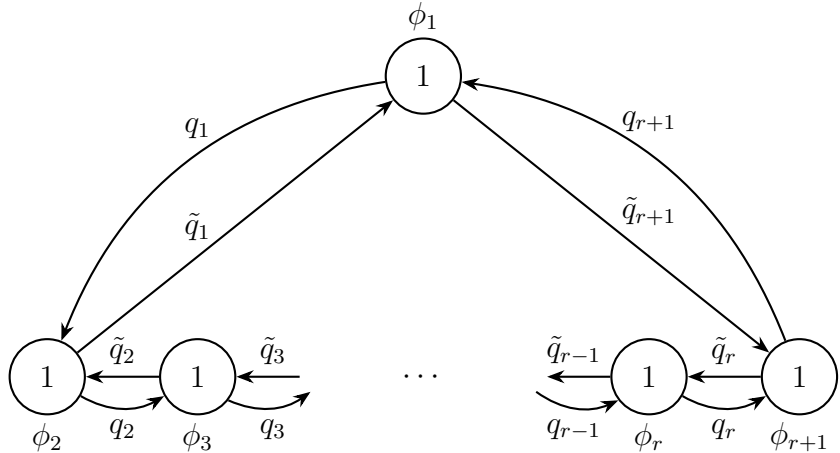

The quiver has the shape of the affine Dynkin diagram of the corresponding ADE Lie algebra \cite{Douglas:1996sw}. 
The nodes of the quiver
correspond to a fractional D2-branes; each node has its  $\mathcal{N}=4$ vector multiplet, with  gauge group  of type $U(n_i)$, where $n_i$ is the dual Coxeter label of the $i$-th node in the associated Dynkin diagram.
Bifundamental hypermultiplets are depicted in the quiver by a pair of arrows connecting the nodes. As an example, Figure~\ref{Fig:ArQuiver} illustrates the $A_r$ quiver.  
In this case, the gauge group is~$\left(\prod_{i=0}^r U(1)\right)/U(1)$. The vector multiplet chiral fields $\phi_i$ are coupled to the hypermultiplets  $(q_i,\tilde{q}_i)$ via the $\mathcal{N}=4$ superpotential
\begin{equation}
W=\sum_{i=1}^{r+1} (\phi_i-\phi_{i+1})q_i\tilde{q}_i, \:,
\end{equation}
with $\phi_{n+2}\equiv \phi_1$. There is also a D-term potential that produces the D-term relations
\begin{equation}\label{eq:D-termAr}
 |q_i|^2+ |\tilde{q}_{i-1}|^2 - |\tilde{q}_{i}|^2 - |q_{i-1}|^2= 0 \qquad\qquad i=1,...,r \:.
\end{equation}

The $A_r$ theory features a single complex mass parameter that cannot be reabsorbed by shifting the fields $\phi_i$; this parameter can be interpreted as part of a background vector multiplet for a $U(1)$ flavor group \cite{Seiberg:1996bs}; the third real scalar is a real mass term (that is allowed in 3d theories). Additionally, the theory possesses $r$ triplets of FI-parameters: the triplet can be separated
in a complex FI-parameter that appears in the superpotential, and a real one that modifies the relations \eqref{eq:D-termAr}; these can be regarded as mass parameters (i.e. background vector multiplets) for an extra rank-$r$ flavor symmetry that is characteristic of 3d theories. This is known as the topological symmetry: for each $U(1)$ factor in the gauge group, there exists an associated conserved $U(1)_T$ current $J_T=\ast F$. This abelian symmetry may enhance to a non-abelian symmetry \cite{Intriligator:1996ex,Gaiotto:2008sa,Bashkirov:2010kz}. In quiver gauge theories, this enhancement occurs in the presence of balanced nodes (namely, when the number of flavors coupled to a gauge node is twice its rank). In the quiver gauge theories arising on the worldvolume of a D2-brane probing an ADE singularity, the topological symmetry enhances to the Lie group generated by the corresponding ADE algebra \cite{Gaiotto:2008sa,Bashkirov:2010kz,Cremonesi:2013lqa}.

The operators charged under the topological symmetry are the {\it monopole operators}, which are local disorder operators\footnote{When inserted in the path integral, a monopole operator induces monopole-like boundary conditions for the magnetic field, with a magnetic flux through a sphere surrounding its insertion point.} that can be directly defined in the infrared conformal field theory  \cite{Borokhov:2002ib}. Given an abelian gauge group $U(1)^n$, there is an associated topological symmetry $U(1)_T^n$, and the monopole operators are labeled by their charges:~$V_{m_1,...,m_n}$. 

For a generic non-abelian gauge group $G$ of rank $\mathsf{r}$, abelian monopole operators can still be constructed from the vector multiplets in the Cartan subalgebra $\mathfrak{h}$ of its Lie algebra. The monopole operators are then labeled as $V_{m_1,...,m_\mathsf{r}}$. To form gauge-invariant operators, one takes Weyl-invariant combinations of  $V_{m_1,...,m_\mathsf{r}}$ along with the complex scalars from the $\mathsf{r}$ vector multiplets.

\subsubsection*{Moduli space}

The moduli space of vacua is a product of two branches, the Higgs branch (HB) and the Coulomb branch (CB).
\begin{itemize}

\item The HB is classically exact and it is parametrized by the vev's of gauge invariant combinations of the hypermultiplet scalars (modulo classical F-term relations). 
They correspond to closed paths in the quiver. F-term equations translate into relations among these operators so that, for the ADE quivers, eventually the dimension of the HB as a complex variety is always 2. In fact, this complex surface reproduces the ADE surface that is probed by the D2-brane.
When the hypermultiplets are given a non-zero vev, the gauge group is fully broken (except for a diagonal $U(1)$, which always decouples). From the brane perspective, this corresponds to merging all the fractional branes into a bound state, now positioned away from the singularity. This is the D2-brane probing the smooth regions of the surface; the decoupled $U(1)$ represents the free vector that propagates along a single D-brane. The scalars within this vector multiplet describe the motion of the brane in the three directions transverse to the ADE surface.

\item The Coulomb branch is sensitive to quantum corrections. At a generic point of this branch, the three scalars in the $\mathcal{N}=4$ vector multiplets develop a vacuum expectation value, leaving massless abelian gauge fields. These massless fields can then be dualized into scalars, and the moduli space becomes a HyperKähler manifold, with its metric subject to quantum corrections. 

The $\mathcal{N}=4$ vector multiplet decomposes into an $\mathcal{N}=2$ vector multiplet and a chiral multiplet transforming in the adjoint representation of the gauge group. In the quantum description of the Coulomb Branch, the $\mathcal{N}=2$ vector multiplets are replaced by monopole operators \cite{Aharony:1997bx}. 
The vev of these fields are subject to some relations. For example, for a 3d $\mathcal{N}=4$ $U(1)$ gauge theory with $N_f$ flavors (i.e. two hypermultiplets with charge 1), there are two monopole operators $V_\pm$ subject to the relation
\begin{equation}
V_+V_- = \prod_{i=1}^{N_f} M_i(\varphi,\mathrm{m}_j)\:,
\end{equation}
with $M_i(\varphi,\mathrm{m}_j)$ the mass of the $i$-th hypermultiplet, as function of the mass parameters $\mathrm{m}_j$ and the vev of the complex scalar $\varphi$ in the $\mathcal{N}=4$ vector multiplet. When $\mathrm{m}_j=0$ $\forall j$, the equation becomes $V_+V_-=\varphi^{N_f}$, i.e. the CB is a surface with an $A_{N_f-1}$ singularity. For higher rank theories the CB geometry is more complicated. 

\end{itemize}

Masses and FI-parameters modify the moduli space, lifting some directions and deforming its geometry. We have just seen the example of a $U(1)$ gauge theory with $N_f$ flavors: switching on complex mass parameters for the hypermultiplets deforms the CB to a smooth space; in addition the parameters $\mathrm{m}_j$ give mass to the hypermultiplets, lifting  the HB. In ADE quiver gauge theories, turning on the FI-parameter associated with a $U(1)$ node means blowing up the corresponding simple root in the singular HB (it corresponds to resolving the surface if one turns on a real FI-parameter, while it corresponds to a deformation for complex FI-parameters). 
This can be understood by recognizing that the FI-parameters are scalars in the background vector multiplets for the topological symmetry and that, in the string theory realization, these background vector multiplets originate from the supergravity $C_3$ form potential and from the geometric moduli.

\subsubsection*{Mirror symmetry}

The HB and the CB are both HyperK\"ahler manifolds. 
The fact that they have the same geometrical structure suggests the existence of theories with the same moduli space where the roles of HB and CB are swapped. Namely, given a theory A with $\mathcal{M}_A=HB_A\times CB_A$, there exists a theory B with $\mathcal{M}_B=HB_B\times CB_B$, such that: 
\be HB_A=CB_B, \quad CB_A=HB_B\:.\ee
Mirror symmetry is the duality that maps $\mathcal{M}_A$ into $\mathcal{M}_B$ \cite{Intriligator:1996ex}. Besides acting on operators, the map sends FI terms into masses and {\it viceversa}. 
The advantages of exploiting mirror symmetry are plural. For example, 
it transforms the problem of determining the quantum relations of the Coulomb branch into an analysis of the Higgs branch, which relies solely on classical considerations.

For the special case of ADE quivers, one can easily reconstruct the mirror dual theories by uplifting the type IIA picture to M-theory and performing the 9-11 flip, mentioned in the previous section. Let us consider the $A_r$ case: the D2-brane probing a singular surface $S_{\rm sing}$ is mapped to a D2-brane probing a stack of $r+1$ D6-branes, that supports a 3d $U(1)$ gauge theory with $r+1$ flavors. 
The mass parameter of the $A_r$ theory is mapped to the single FI-parameter of the $U(1)$ theory. The masses $\mathrm{m}_j$ of the $U(1)$ theory are vev's of scalars in the  background $SU(r+1)$ vector multiplets (which live on the D6-brane stack and describe their motion in transverse directions); these are mapped to the vector multiplets coming from reducing the $C_3$ 3-form on the two-forms of the surface $S$ (the scalars are geometric moduli controlling the deformation and the resolution of the singularity). The corresponding bulk $U(1)^r$ symmetry is the topological symmetry for the probe 3d theory.

\subsection{D2-branes probing an $A_1$ singularity and its mirror dual}\label{Sec:A1SingAnd3dMirro}

We now focus on the simplest case, the $A_1$ singularity, as a concrete example.  Not only does this case illustrate the key ideas, but it also serves as the building block for analyzing more complicated singularities.

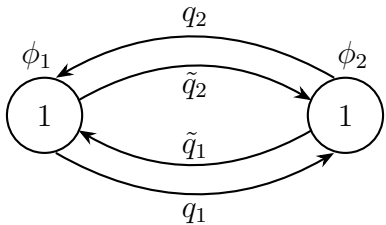
\begin{figure}[H]
\captionsetup{justification=centering}
 \centering
\begin{tikzpicture} [place/.style={circle,draw=black!500,fill=white!20,thick,
inner sep=0pt,minimum size=10mm}, transition/.style={rectangle,draw=black!50,fill=black!20,thick,
inner sep=0pt,minimum size=10mm}, decoration={ markings,
mark=between positions 0.35 and 0.85 step 2mm with {\arrow{stealth}}}]
  \node at ( 2,0) [place] (0) {$1$};  \node at ( 2.1,0.8)  {$\phi_2$}; 
  \node at (- 2,0) [place] (1) {$1$};  \node at ( -2.1,0.8)  {$\phi_1$}; 
\draw[thick, -Stealth] (-1.55,0.2)[bend left] to (1.55,0.2); \node at ( 0,1.3)  {$q_2$}; 
\draw[thick, Stealth-] (-1.55,-0.2)[bend right] to (1.55,-0.2);\node at ( 0,-1.3)  {$q_1$};  
\draw[thick, Stealth-] (-1.85,0.5)[bend left] to(1.85,0.5);\node at ( 0,0.4)  {$\tilde{q}_2$}; 
\draw[thick, Stealth-] (1.85,-0.5)[bend left] to(-1.85,-0.5);\node at ( 0,-0.4)  {$\tilde{q}_1$}; 
\end{tikzpicture}
\caption{$A_1$ quiver.}\label{Fig:A1quiver}
\end{figure}

The worldvolume theory of a D2-brane probing the $A_1$ singularity is a 3d $\mathcal{N}=4$  supersymmetric 	quiver gauge theory with the quiver in Figure~\ref{Fig:A1quiver}  and the superpotential
\begin{equation}
 W_B = (\phi_1-\phi_2)(q_1\tilde{q}_1-q_2\tilde{q}_2)\:.
\end{equation}
We call this `Theory B'. We note that one combination of the two $U(1)$'s decouples, leaving a $U(1)$ gauge theory with $N_f=2$ charged hypermultiplets; the scalar field in the coupled $U(1)$ vector multiplet is $\phi_-\equiv \phi_1-\phi_2$. This theory is self-mirror; hence, applying mirror symmetry one obtains again a $U(1)$ gauge theory with two hypermultiplets that are charged under the $U(1)$ gauge group with the following superpotential (plus a decoupled hypermultiplet):
\begin{equation}
W_A = \varphi (Q_1\tilde{Q}_1+Q_2\tilde{Q}_2)\:.
\end{equation}
We call this `Theory A'. This is the worldvolume theory of a D2-brane probing a stack of two D6-branes living in flat space-time in type IIA, that by 9-11 flip is dual to the background probed in Theory B.

The mirror symmetry maps gauge invariant operators of Theory B to gauge invariant operators of Theory A and {\it viceversa}. For  $\mathcal{N}=4$ 3d theories it in particular exchanges the Higgs Branch with the Coulomb branch.

In this simple example the mirror map is explicitly known (see e.g. \cite{Aharony:1997bx}). When it acts on CB$_B$ and  HB$_A$ (exchanging them), one has\footnote{To be precise, the map works once one imposes the F-term of $\varphi$ on the A-side. In all the cases we consider in this paper this holds.}
\begin{equation}\label{eq:MirrMapCBBHBA}
\begin{pmatrix}
\phi_- & w_+ \\ w_- & -\phi_- \\
\end{pmatrix}                  \leftrightarrow 
\begin{pmatrix}
Q_1\tilde{Q}_1 & Q_1\tilde{Q}_2 \\ Q_2\tilde{Q}_1  & Q_2\tilde{Q}_2 
\end{pmatrix}\:,
\end{equation}
where $w_\pm$ are the monopole operators associated with the relative U(1) in Theory B and $M_{ij}=Q_i\tilde{Q}_j$ is the meson matrix of Theory A. These are the gauge invariant coordinates of respectively CB$_B$ and HB$_A$.

The map exchanging HB$_B$ $\leftrightarrow$ CB$_A$ is:\footnote{As before,  the map works once one imposes the F-term of $\phi_-$ on the B-side. In all the cases we consider in this paper this relation is satisfied.}
\begin{equation}\label{eq:MirrMapHBBCBA}
\begin{pmatrix}
q_1\tilde{q}_1 & -q_1 q_2 \\ \tilde{q}_2 \tilde{q_1}  & -\tilde{q}_2 q_2 
\end{pmatrix}
\leftrightarrow 
\begin{pmatrix}
\varphi & -v_+ \\ v_- & -\varphi \\
\end{pmatrix}                  
\end{equation}
where $q_i\tilde{q}_j$ are the gauge invariant coordinates on HB$_B$ and $\varphi,v_\pm$ are the gauge invariant coordinates on CB$_A$.

\section{D2-branes and families of ADE surfaces}\label{Sec:D2deformedADEfamilies}

We now switch on a vev for $\Phi$. This will deform the worldvolume theory of the D2-brane as it is now probing a different background (deformed by non-zero $\Phi$).

\subsection{$A_1$-family from D2-branes and universal flop of length 1}\label{Sec:A1famFromD2}

We start again considering the $A_1$ case studied in Section~\ref{Sec:A1SingAnd3dMirro}.

On the Theory A side, the effect of a non-zero vev for $\Phi$ is straightforward to understand, as $\Phi$ is the complex adjoint scalar living on the worldvolume of a stack of two D6-branes. This induces the following superpotential deformation on the worldvolume of the probing D2-brane:
\begin{equation}\label{eq:SupDefThAinA1}
\delta W_A = \tr \Phi M \qquad\mbox{with}\qquad {M_{ij}}=Q_i\tilde{Q}_j \:,
\end{equation}
that is a mass term for some flavors. 

Applying the mirror  maps to \eqref{eq:SupDefThAinA1}, one obtains the superpotential deformation induced by $\Phi$ in Theory B.
We notice that when the matrix $\Phi$ has off-diagonal elements, the trace in \eqref{eq:SupDefThAinA1} involves operators $Q_i\tilde{Q}_j$ with $i\neq j$ that  are mapped by mirror symmetry to monopole operators (see \eqref{eq:MirrMapCBBHBA}). Hence in this case $\delta W_B$  includes these non elementary operators. In order to deal with monopole superpotentials  we follow the procedure in \cite{Collinucci:2016hpz,Collinucci:2017bwv,Benini:2017dud}: one maps the 3d theory with monopole superpotential to a dual 3d theory with a superpotential that depends only on elementary fields. 

We consider again the $A_1$ case to illustrate the logic we follow in this paper. We work in Theory B.
In the $A_1$ example, a non-zero vev for $\Phi$ generates  the deformation
\begin{equation}\label{eq:WBA1case}
\delta W_B = \tr \left[ \Phi(\varrho) \, \begin{pmatrix}
\phi_1-\phi_2 & w_+ \\ w_- & \phi_2-\phi_1 \\
\end{pmatrix}   \right] \:.
\end{equation}
We analyze both choices of $\Phi$ in the equation \eqref{eq:PhiA12choices}: the first one is the $\Phi$ corresponding to the universal flop of length one, the second one will be important for addressing the more complicated cases discussed in the upcoming sections.
\begin{itemize}
\item Case 1: $$\delta W_B = 2 t (\phi_1-\phi_2).$$ This is a complex FI term. The Higgs branch of the moduli space is simply deformed to 
$$uv=z^2-t^2$$
in terms of the gauge invariant variables $u=q_1q_2$, $v=\tilde{q}_1\tilde{q}_2$ and $z=q_1\tilde{q}_1+t$. This is the expression for the $A_1$ family with simultaneous resolution, i.e. {\it  the universal flop of length one}. 

Let us check the presence of other branches allowed by the F-terms: When $t\neq 0$ the $U(1)$ gauge group of the 3d theory is broken and the HB is the only branch of the theory. At $t=0$, a new branch arises, i.e. the CB with geometry $\mathbb{C}^2/\mathbb{Z}_2$; moreover, along this branch the $U(1)$ gauge group is unbroken. 
\item Case 2: 
\begin{equation}\label{eq:WsupdefThBA1case} 
\delta W_B = w_- + \mu \, w_+.
\end{equation}
This superpotential depends on monopole operators. We go to the mirror theory, where the deformed superpotential is 
\begin{equation}\label{eq:WsupdefThAA1case} 
W_A^{\rm def} = \varphi (Q_1\tilde{Q}_1+Q_2\tilde{Q}_2) + Q_2\tilde{Q}_1+ \mu Q_1\tilde{Q}_2
\end{equation}
The deformations are now simple mass terms. Since we want to obtain the moduli space for each value of the parameter $\mu$, we integrate only the fields that are massive $\forall \mu$, i.e. in this case $Q_2$ and $\tilde{Q}_1$. After doing this, one obtains an $\mathcal{N}=2$ 3d\footnote{When $\mu\neq 0$, the 3d $\mathcal{N}=4$ supersymmetry is recovered as emergent in the IR \cite{Carta:2018qke,Maruyoshi:2016tqk,Maruyoshi:2016aim}. In fact, \cite{Carta:2018qke} considered deformation like the one in this example, with $\mu$ a dynamical field.} U(1) gauge theory with one flavor, one singlet $\varphi$ and the superpotential
\begin{equation}\label{eq:WAeffA1}
W_{A,\rm eff}= (\mu-\varphi^2)Q_1\tilde{Q}_2\:.
\end{equation}
We now apply mirror symmetry on this effective theory \cite{Collinucci:2016hpz,Collinucci:2017bwv}. The mirror symmetric of a U(1) gauge theory with one flavor and zero superpotential is the $XYZ$ model, with mirror map given by $X\leftrightarrow Q_1\tilde{Q}_2$, $Y\leftrightarrow v_+$, $Z\leftrightarrow v_-$. When the U(1) model is deformed by the superpotential \eqref{eq:WAeffA1}, the $XYZ$ model has the superpotential
\begin{equation}
W_{B,\rm eff} = X Y Z + X (\mu-\varphi^2) = X\,\det \mathfrak{M} + \mu \, X\,, \qquad \mbox{with } \mathfrak{M}\equiv \begin{pmatrix}
\varphi & -Y \\ Z & -\varphi \\
\end{pmatrix}\:.
\end{equation}
The matrix $\mathfrak{M}$ is in the adjoint representation\footnote{One can interpret the fields $\varphi,Y,Z$ as condensates of $q_i,\tilde{q_j}$ (see \eqref{eq:MirrMapHBBCBA}).} of the original flavor group $SU(2)$ (that was acting on the HB) that is not broken by the deformation $\delta W_B$. 
The F-terms are
\begin{eqnarray}
 \frac{\partial W}{\partial X}=0 &  : \qquad & Y \, Z =\varphi^2-\mu \:; \label{eq:FtermA1eff1}\\
 \frac{\partial W}{\partial \varphi}=0 & : \qquad  & X\, \varphi  = 0 \:;\\ 
 \frac{\partial W}{\partial Y}=0 & : \qquad  & X\, Z =0 \:;\\
 \frac{\partial W}{\partial Z}=0 & : \qquad  & X\, Y =0  \:.
\end{eqnarray}
The last two equations give two branches: $X=0$ or $Y=Z=\varphi=0$. The second one is excluded when $\mu\neq0$, because of \eqref{eq:FtermA1eff1}. The first one is the space parametrized by $Y,Z,\varphi$ subject to the relation $YZ=\varphi^2-\mu$, that is the defining equation of the $A_1$ family with no simultaneous resolution. 
For $\mu=0$, the branch $Y=Z=\varphi=0$ is allowed. It is a complex one-dimensional space parametrized by $X$. There is still no $U(1)$ gauge group when $\mu=0$.
\end{itemize}

From the point of view of the probe D2-brane, the two $\Phi$'s in \eqref{eq:PhiA12choices} deform the worldvolume theory in the same manner  away from the origin ($t\neq 0$ or $\mu\neq 0$); at this point however the worldvolume theory of the probe D2-brane is different even though the geometry is the singular $A_1$ surface. What distinguishes the two $\Phi$'s at the origin is that in Case~2, the geometry is supplemented by a T-brane background. The theory on a D2-brane probing a T-brane background was studied first in \cite{Collinucci:2016hpz} and is exactly the one we have just discussed. In particular, we notice that the original CB is partially lifted to the $\mathbb{C}$ parametrized by $X$.

\subsection{ADE family from the D2- probe for generic ADE algebra}\label{Sec:GenericADEalgSuperpotDef}


We now extend the lesson we learned in the simple $A_1$ example to a generic ADE singularity. One probes  such a singularity by a D2-brane. The singular background can be deformed by switching on the adjoint complex scalar $\Phi$. From the point of view of the probe theory, it is an object in the adjoint  representation of the flavor group $G_F$ acting on one branch of the moduli space (the CB on the B-side) of the probe 3d theory. The flavor algebra is the ADE algebra associated with the singularity.

The deformation of the superpotential takes an elementary form:
\begin{equation}\label{eq:supDefGenericADE}
 \delta W = \tr \left[ \Phi \bm{\mu} \right] \:,
\end{equation}
where $\bm{\mu}$ is the moment map of the flavor symmetry $G_F$.
In Theory B (D2-branes at ADE singularities) $\bm{\mu}$ is the moment map on the CB, that involves several monopole operators. Hence, to deal with this deformation one needs to be able to treat monopole superpotentials. We will illustrate how to do this in the following section, in preparation for the more complicated universal flop of length 2.

The form \eqref{eq:supDefGenericADE} of the deformation can be generalized to any ADE algebra. It can be checked explicitly by string duality for the $A$ and $D$ algebras (as done for the $A_1$ case).
Adding \eqref{eq:supDefGenericADE} to the superpotential 
deforms the Higgs branch of the probe 3d theory. The family of such deformed HB moduli spaces over the parameter space $B_\varrho$ {\it is } the ADE-family, with the simultaneous resolution determined by the form of $\Phi$ or, equivalently, by the Levi subalgebra in which $\Phi(\varrho)$ resides.

\

The effective 3d theory provides more information than the defining equation of the ADE family. 
When the parameters $\boldsymbol{\varrho}$ are such that the ADE surface becomes singular, the HB of the D2-brane probe theory develops singularities, and new branches of the moduli space emerge from those singularities. 
However, this does not imply that the entire $(r + 2)$-fold family $X_{r+2}$ becomes singular at that value of $\boldsymbol{\varrho}$. A singularity arises in the family when, upon resolving it, certain roots of the singular fiber blow up.\footnote{This means that actual curves have shrunk to zero size in the singular limit.} In the effective 3d theory, the sizes of these roots are determined by the FI parameters, as explained in Section~\ref{sec:susy3dtheories}.
Activating one of these parameters forces the corresponding root to blow up at the singularity of the HB. These FI parameters correspond one-to-one with the $U(1)$ factors in the gauge group of the 3d theory.
When the effective theory lacks any $U(1)$ factors, even in the presence of a singular HB, it indicates a T-brane background, which obstructs the resolution \cite{Collinucci:2016hpz}. This means that no roots can be resolved at that point in $B_\varrho$, and consequently, the family is not singular there. This is the mechanism by which partial simultaneous resolutions are implemented in the effective theory.

In other words, the deformation \eqref{eq:supDefGenericADE} leads to an effective 3d theory with its own quiver (i.e. gauge group and matter) and superpotential $W_{\rm eff}$; varying $\boldsymbol{\varrho}$ the superpotential changes and the moduli space of the 3d theory is modified. At generic points in $B_\varrho$ the moduli space has only one branch, the deformed HB, where the gauge group is broken completely. At points of $B_\varrho$ where a new branch arises, the effective theory may have points in the moduli space (in particular the common origin of the branches) where (part of) the gauge group of the quiver is unbroken. If this (sub)group contains $U(1)$ factors, there is the possibility to activate real FI-parameters, that lead to the resolution of the singularity of the HB.

We therefore present the following statement, which will be supported by the examples that follow:  

\

\noindent
 {\it  Consider the set  of effective 3d theories over the space  $B_\varrho$. 
The family of HB moduli spaces {\it is} the family of ADE-surfaces.  
Singular surfaces arise at values of $\boldsymbol{\varrho}$ where the 3d theory develops an additional branch beyond the HB. Furthermore, at such points in $B_\varrho$, the gauge group of the effective theory at the intersection of the new branch with the HB may or may not have some $U(1)$ factors; the number of the corresponding FI-parameters indicates how many roots of the singular fiber are blown up in the (partial) simultaneous resolution.
 }

\

\noindent

This statement is, in particular, verified in the $A_1$ case studied above. There, the gauge group has only one $U(1)$ factor, hence there is at most one FI-parameter. This is related  to the fact that $A_1$ has one simple root, that may or may not be resolved in the family, corresponding to the fact that the in the effective theory the $U(1)$ survives or not. Let us see the two possible cases:
\begin{itemize}
\item Case 1. At $t=0$, a new branch arises, corresponding to the fiber developing an $A_1$ singularity, and a $U(1)$ factor survives, with its real FI-term that would blow up the simple root of the $A_1$ singularity. This is in agreement with our statement, as Case 1 is the family {\it with} simultaneous resolution, i.e. the {\it universal flop of length $\ell=1$}.
\item Case 2. When $\mu\neq 0$ the deformed $A_1$ surface is the only branch of the moduli space. 
At $\mu=0$ we have a new branch that tells us the corresponding ADE surface is singular as it can be checked from the defining equation of the family. However, now there is no $U(1)$ factor in the gauge group, hence the effective theory cannot be deformed by a non-zero real FI-term, that means that the shrunk $A_1$ root cannot be blown up. This is compatible with the fact that this family has no simultaneous resolution of the simple root in the central fiber (as this family is a smooth space).
\end{itemize}

Another simple example to test our statement is an $A_2$ family with complete simultaneous resolution. The corresponding Higgs field is a diagonal matrix, $\Phi=$Diag$(t_1,t_2,-t_1-t_2)$, where $t_1,t_2$ are the parameters parametrizing the base $B_\varrho$. The defining equation takes the form
\begin{equation}\label{eq:equationFamilyA2}
u\,v=(z-t_1)(z-t_2)(z+t_1+t_2) \:.
\end{equation}
The superpotential deformation is just a sum of complex FI-term:
\begin{equation}\label{eq:deltaWA2}
\delta W=(2t_1+t_2)\phi_1 - (t_1-t_2)\phi_2 - (2t_2+t_1)\phi_3 \:.
\end{equation}
At a generic point in $B_\varrho$, the $U(1) \times U(1)$ gauge symmetry is fully broken, and the HB is deformed into a smooth surface, which remains the only branch of the moduli space. When the coefficient of one of the $\phi_i$ in \eqref{eq:deltaWA2} vanishes, the corresponding $U(1)$ factor is preserved in the effective theory, and its CB emanates from the HB. The effective theory possesses a real FI parameter, consistent with the fact that at these points, the equation \eqref{eq:equationFamilyA2} develops an $A_1$ singularity, which is blown up in the simultaneous resolution of the family. Finally, when $t_1 = t_2 = 0$, the gauge group retains both $U(1)$ factors with corresponding FI parameters and CB. This is the origin of the family, where the surface has an $A_2$ singularity, and both roots are blown up in the simultaneous resolution.

\subsection{Example: $A_3$-family from D2-branes}\label{Sec:A3example}

As a preliminary step toward tackling the universal flop of length $\ell=2$, we first consider a simpler~$A_n$~case. 

\begin{figure}[H]
 \centering
\begin{tikzpicture} [place/.style={circle,draw=black!500,fill=white!20,thick,
inner sep=0pt,minimum size=5mm}, transition/.style={rectangle,draw=black!50,fill=black!20,thick,
inner sep=0pt,minimum size=10mm}, decoration={ markings,
mark=between positions 0.35 and 0.85 step 2mm with {\arrow{stealth}}}]
 \node at ( 3,0) [place] (3) {}; \node at ( 3,-0.6) {$1$}; \node at (3,0.6) {$\alpha_3$};
  \node at (- 3,0) [place] (1) {}; \node at (- 3,-0.6) {$1$}; \node at (- 3,0.6) {$\alpha_1$};
 \node at (0,0) [place, fill=black] (2) {};\node at (0,-0.6) {$1$};\node at (0,0.6) {$\alpha_2$};
  \draw[thick] (-2.5,0)--(-0.5,0);
   \draw[thick] (2.5,0)--(0.5,0);
\end{tikzpicture}
\caption{$A_3$ Dynkin diagram. The colored node corresponds to the blown up sphere in the partial simultaneous resolution of the $A_3$ family.}\label{Fig:A3partialRes}
\end{figure}
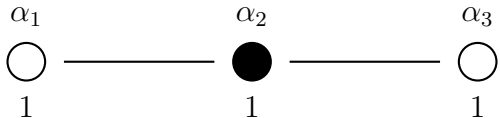
We want to realize a family of $A_3$ surfaces with the simultaneous resolution only of the central node of the corresponding Dynkin diagram (see Figure~\ref{Fig:A3partialRes}). This means that the Levi subalgebra $\Phi(\varrho)$ should belong to is 
\begin{equation}
 \mathcal{L} = A_1^{\alpha_1} \oplus A_1^{\alpha_3} \oplus \langle \alpha_2^* \rangle
\end{equation}
where $A_1^{\alpha_i}$ is the $A_1$ subalgebra $\langle e_{\alpha_i},e_{-\alpha_i},[e_{\alpha_i},e_{-\alpha_i}]\rangle$.

Following the rules given in Section~\ref{sec:FamilyFromHiggs}, the corresponding Higgs field takes the form
\begin{equation}\label{eq:PhiA3simple}
 \Phi = \begin{pmatrix}
  	\varrho_2 & 1 & & \\ \varrho_1 &  \varrho_2 & & \\ & & -\varrho_2 & 1 \\ & & \varrho_3 & - \varrho_2 \\
 \end{pmatrix}
\end{equation}
and the equation for the family is, according to \eqref{Eq:DefSAn},
\begin{equation}\label{eq:A3family}
uv=\left((z-\varrho_2)^2-\varrho_1 \right) \left((z+\varrho_2)^2-\varrho_3 \right) \:.
\end{equation}

We now consider a D2-brane probing the $A_3$ surface deformed by \eqref{eq:PhiA3simple} and we will show that the family of its HB moduli spaces over $B_\varrho$ is exactly the family \eqref{eq:A3family}.

\begin{figure}[t!]
\captionsetup{justification=centering}
 \centering
\begin{tikzpicture} [place/.style={circle,draw=black!500,fill=white!20,thick,
inner sep=0pt,minimum size=10mm}, transition/.style={rectangle,draw=black!50,fill=black!20,thick,
inner sep=0pt,minimum size=10mm}, decoration={ markings,
mark=between positions 0.35 and 0.85 step 2mm with {\arrow{stealth}}}]
  \node at ( 3,0) [place] (3) {$1$}; \node at ( 3,-0.8)  {$\phi_4$};  \node at ( 1.5,-0.8)  {$q_3$};  \node at ( 1.5,0.3)  {$\tilde{q}_3$}; 
  \node at (- 3,0) [place] (1) {$1$};  \node at ( -3,-0.8)  {$\phi_2$};  \node at ( -1.5,-0.8)  {$q_2$};  \node at ( -1.5,0.3)  {$\tilde{q}_2$}; 
  \node at (0,2.5)[place] (0) {$1$}; \node at (0, 3.3) {$\phi_1$};  \node at ( -1.5,2.3)  {$q_1$};  \node at ( -1.5,1.6)  {$\tilde{q}_1$}; 
     \node at (0,0)[place]  (2) {$1$};  \node at ( 0,-0.8)  {$\phi_3$};  \node at ( 1.5,2.3)  {$q_4$};  \node at ( 1.5,1.6)  {$\tilde{q}_4$}; 
 \draw[thick, -Stealth] (0)[bend right] to(1);\draw[thick, -Stealth] (1)--(0);
 \draw[thick, Stealth-] (0)[bend left] to (3);\draw[thick, Stealth-] (3)--(0);
 \draw[thick, -Stealth] (2)[bend right] to (3);\draw[thick, -Stealth] (3)--(2);
 \draw[thick, -Stealth] (1)[bend right] to (2);\draw[thick, -Stealth] (2)--(1);
\end{tikzpicture}
\caption{$A_3$ quiver.}\label{Fig:A3quiver}
\end{figure}
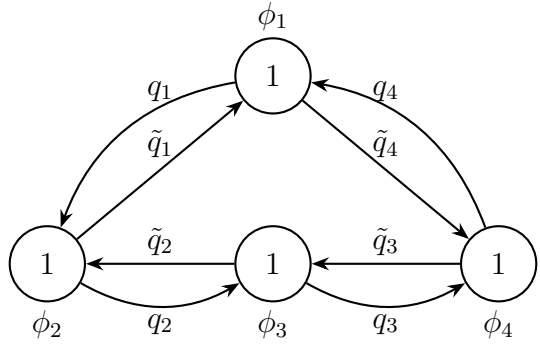

The worldvolume theory of a D2-brane probing the $A_3$ singularity is a 3d $\mathcal{N}=4$ supersymmetric quiver gauge theory, with the quiver in Figure~\ref{Fig:A3quiver} and the superpotential
\begin{equation}
W_{A_3}=\sum_{i=1}^4 (\phi_i-\phi_{i+1})q_i\tilde{q}_i \:,
\end{equation}
with the convention $\phi_{4+1}=\phi_1$.

To write down the deformation \eqref{eq:supDefGenericADE} that $\Phi$ induces, we need to say what is the moment map on the CB of this theory. The flavor symmetry on the CB is $SU(4)$ and the moment map is (according to mirror symmetry \cite{Intriligator:1996ex,Aharony:1997bx})
\begin{equation}\label{eq:momentMapA3CB}
\bm{\mu} = \begin{pmatrix}
\phi_1-\phi_2 & w_{1,0,0} & w_{1,1,0} & w_{1,1,1} \\
w_{-1,0,0} & \phi_2-\phi_3 &   w_{0,1,0}  & w_{0,1,1} \\
w_{-1,-1,0} & w_{0,-1,0} &  \phi_3-\phi_4 &    w_{0,0,1} \\
w_{-1,-1,-1} & w_{0,-1,-1} & w_{0,0,-1} &  \phi_4-\phi_1 & \\
\end{pmatrix}\:.
\end{equation}
Plugging \eqref{eq:PhiA3simple} and \eqref{eq:momentMapA3CB} into $\delta W = \tr\left[ \Phi \,\bm{\mu}\right]$, we obtain
\begin{equation}
\delta W = 2 \varrho_2(\phi_1-\phi_3)+w_{-1,0,0}+\varrho_1 w_{1,0,0}+w_{0,0,-1}+\varrho_3 w_{0,0,1} \:.
\end{equation}
The monopole operators that appear in this deformation are those related to the white nodes (2 and 4), i.e. the roots that are not blown up in the partial simultaneous resolution of the family.

We now write the deformed superpotential $ W^{\rm def} = W_{A_3} + \delta W$ in the following form
\begin{eqnarray}\label{eq:FullDefSuperpotA3case}
 W^{\rm def} &=& \phi_2 \tr \mathcal{M}_2 +w_{-1,0,0}+\varrho_1 w_{1,0,0} + \tr \left[ \begin{pmatrix}
  \phi_1 &  0 \\ 0 & \phi_3  
 \end{pmatrix} \mathcal{M}_2 \right]  \nonumber\\
&+&  2 \varrho_2(\phi_1-\phi_3) \\
&+&\phi_4 \tr \mathcal{M}_4  +w_{0,0,-1}+\varrho_3 w_{0,0,1} + \tr \left[\begin{pmatrix}
  \phi_3 &  0 \\ 0 & \phi_1  
 \end{pmatrix} \mathcal{M}_4 \right]\:,\nonumber
 \end{eqnarray}
where $\mathcal{M}_i$, $i=2,4$, are the meson matrices of  nodes 2 and 4, i.e.
\begin{equation}
\mathcal{M}_2 = \begin{pmatrix}
q_1 \tilde{q}_1 & -q_1q_2 \\ \tilde{q}_2\tilde{q}_1 & -\tilde{q}_2 q_2 \\
\end{pmatrix} \qquad \mbox{and} \qquad 
\mathcal{M}_4 = \begin{pmatrix}
q_3 \tilde{q}_3 & -q_3q_4 \\ \tilde{q}_4\tilde{q}_3 & -\tilde{q}_4 q_4 \\
\end{pmatrix}\:.
\end{equation}

The superpotential operators in \eqref{eq:FullDefSuperpotA3case} are relative to the U(1) gauge group of nodes 2 and 4 respectively. In order to deal with such situations, with monopole operators in the superpotential, \cite{Collinucci:2016hpz} proposed the following procedure: 
\begin{figure}[H]
 \centering
\begin{tikzpicture} [place/.style={circle,draw=black!500,fill=white!20,thick,
inner sep=0pt,minimum size=10mm}, transition/.style={rectangle,draw=black!500,fill=white!20,thick,
inner sep=0pt,minimum size=10mm}, decoration={ markings,
mark=between positions 0.35 and 0.85 step 2mm with {\arrow{stealth}}}]
  \node at ( 3,0) [place] (3) {$1$}; 
  \node at (- 3,0) [place] (1) {$1$}; 
  \node at (0,2.5)[transition] (0) {$1$}; 
     \node at (0,0)[transition]  (2) {$1$}; 
      \node at ( 3,-0.8)  {$\phi_4$};  \node at ( 1.5,-0.8)  {$q_3$};  \node at ( 1.5,0.3)  {$\tilde{q}_3$}; 
 \node at ( -3,-0.8)  {$\phi_2$};  \node at ( -1.5,-0.8)  {$q_2$};  \node at ( -1.5,0.3)  {$\tilde{q}_2$}; 
\node at (0, 3.3) {$\phi_1$};  \node at ( -1.5,2.3)  {$q_1$};  \node at ( -1.5,1.6)  {$\tilde{q}_1$}; 
 \node at ( 0,-0.8)  {$\phi_3$};  \node at ( 1.5,2.3)  {$q_4$};  \node at ( 1.5,1.6)  {$\tilde{q}_4$}; 
 \draw[thick, -Stealth, red] (0)[bend right] to(1);\draw[thick, -Stealth, red] (1)--(0);
 \draw[thick, Stealth-, blue] (0)[bend left] to (3);\draw[thick, Stealth-, blue] (3)--(0);
 \draw[thick, -Stealth, blue] (2)[bend right] to (3);\draw[thick, -Stealth, blue] (3)--(2);
 \draw[thick, -Stealth, red] (1)[bend right] to (2);\draw[thick, -Stealth, red] (2)--(1);
\draw[->, thick] (4, 1)--(5.5, 1); 
  \node at (7,0)[place]  (4) {$1$}; \node at (7.1, 0.8) {$\phi_4$}; \node at (8.5, -0.9) {$(q_3, \tilde{q}_4)$};\node at (8.5, 0.9) {$(q_4, \tilde{q}_3)$};
   \node at (10,0)[transition, fill=blue!60]  (5) {$2$}; \draw[thick, Stealth-, blue] (7.4,0.3)[bend left] to (9.5,0.3); \draw[thick, -Stealth, blue] (7.4,-0.3)[bend right] to (9.5,-0.3);
    \node at (7,3.2)[place]  (6) {$1$}; \node at (7.1, 4) {$\phi_2$};\node at (8.5, 2.3) {$(q_2, \tilde{q}_1)$};\node at (8.5, 4.1) {$(q_1, \tilde{q}_2)$};
 \node at (10,3.2)[transition, fill=red!70]  (7) {$2$}; \draw[thick, Stealth-, red] (7.4,3.5)[bend left] to (9.5,3.5); \draw[thick, -Stealth, red] (7.4,2.9)[bend right] to (9.5,2.9);   
\end{tikzpicture}
\caption{$A_3$ theory: ungauging nodes 1 and 3, we end up with two copies of a  $U(1)$ gauge theory with 2 flavors.}\label{Fig:localMirror}
\end{figure}
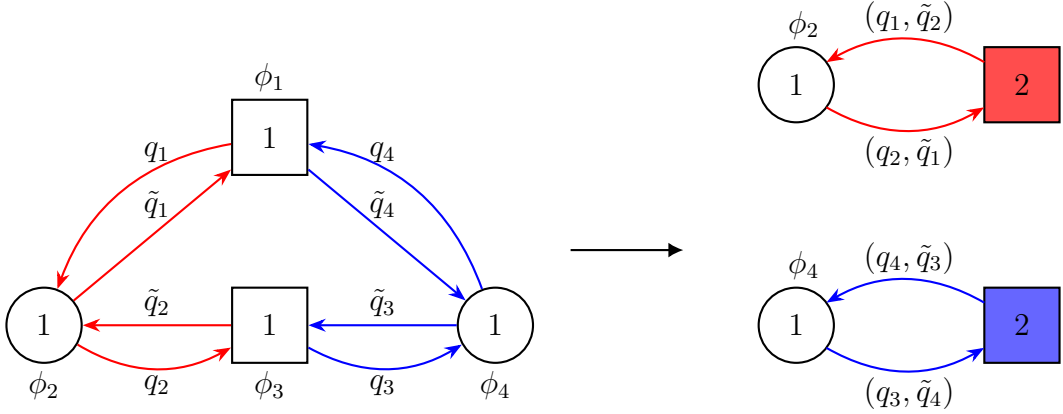
\begin{enumerate}
\item The monopole operator is charged under the topological $U(1)_T$ relative to one node. Consider that abelian node, and ungauge the nearby U(1) gauge factors; in our example, one ungauges node 1 and 3, isolating nodes 2 and 4 as depicted in Figure~\ref{Fig:localMirror}. The isolated theory is coupled back to the nearby nodes  (to obtain the full quiver gauge theory) by gauging a subgroup of its flavor symmetry; this is implemented by the last terms in row 1 and~3~in~\eqref{eq:FullDefSuperpotA3case}. 

For each isolated (balanced) node one has a 3d $\mathcal{N}=4$ supersymmetric $U(1)$ gauge theory with two flavors and a monopole superpotential deformation.
Let us consider node 2 in our example: The isolated theory is $U(1)$ with two flavors and superpotential
\begin{equation}
W_2 =  \phi_2 \tr \mathcal{M}_2 +w_{0,-1,0,0}+\varrho_1 w_{0,1,0,0} \:.
\end{equation}
This is clearly of the form \eqref{eq:WBA1case} encountered in the $A_1$ case, with $\mu=\varrho_1$. We have the same for node 4, with $\mu=\varrho_3$.

\item Now, we apply (`local') mirror symmetry to the isolated 3d theory to go to a $U(1)$ gauge theory with two flavors and mass term deformations like in the $A_1$ case. We then compute the effective theory after integrating out the massive field (massive for all values of $\boldsymbol{\varrho}$). Finally, we apply mirror symmetry back obtaining an effective theory. This theory has the same flavor symmetry as the original isolated theory on its HB. 

Basically we need to follow the same steps as in the $A_1$ case of Section~\ref{Sec:A1famFromD2}. We obtain a modified XYZ theory with superpotential 
\begin{equation}
W_2^{\rm eff} = X_2 \det \mathfrak{M}_2 + \varrho_1 \, X_2 \qquad\mbox{with}\qquad 
\mathfrak{M}_2 \equiv \begin{pmatrix}
 \varphi_2 & -Y_2 \\ Z_2 & -\varphi_2 \\
\end{pmatrix}\:
\end{equation}
and flavor symmetry $SU(2)$. For node 4 we have the same theory with analogous superpotential.

\item We now have to couple this effective theory back to the quiver. The superpotential terms realizing it are traces of the product of the moment map of the $SU(2)$ flavor symmetry with the $\phi$-dependent matrix appearing at the end of rows 1 and 3 in \eqref{eq:FullDefSuperpotA3case}.

For node 2, we need to add 
\begin{equation}
 \tr \left[ \begin{pmatrix}
  \phi_1 & 0 \\ 0 & \phi_3 \\
 \end{pmatrix} \mathfrak{M}_2 \right]= 
  \tr \left[ \begin{pmatrix}
  \frac{\phi_1-\phi_3}{2} & 0 \\ 0 & \frac{\phi_3-\phi_1}{2} \\
 \end{pmatrix} \mathfrak{M}_2\right] = 
 \varphi_2(\phi_1-\phi_3) \:.
\end{equation}
We do analogously for node 4.
\end{enumerate}

We can now write down the effective superpotential for the effective theory, by 
adding also the term in the second line of \eqref{eq:FullDefSuperpotA3case}:
\begin{equation}
 W^{\rm eff} = X_2\left( \det\mathfrak{M}_2 + \varrho_1\right) + X_4 \left( \det\mathfrak{M}_4 + \varrho_3\right) + (\varphi_2-\varphi_4+2\varrho_2)(\phi_1-\phi_3)  \:.
\end{equation}

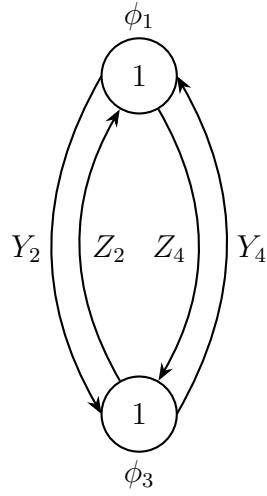
\begin{figure}[t]
\captionsetup{justification=centering}
 \centering
\scalebox{1}{ 
\begin{tikzpicture} [place/.style={circle,draw=black!500,fill=white!20,thick,
inner sep=0pt,minimum size=10mm}, transition/.style={rectangle,draw=black!50,fill=black!20,thick,
inner sep=0pt,minimum size=10mm}, decoration={ markings,
mark=between positions 0.35 and 0.85 step 2mm with {\arrow{stealth}}}]
  \node at (0,4.5)[place] (0) {$1$}; \node at (0, 5.3) {$\phi_1$};
  \node at (-1.5, 2.2) {$Y_2$};
   \node at (-0.4, 2.2) {$Z_2$};
    \node at (1.5, 2.2) {$Y_4$};
     \node at (0.4, 2.2) {$Z_4$};
     \node at (0,0)[place]  (2) {$1$}; \node at (0, -0.8) {$\phi_3$};
    \draw[thick, -Stealth]  (0)[bend left] to (2);
      \draw[thick, Stealth-]  (0)[bend right] to (2);
 \draw[thick, -Stealth]  (-0.5,4.5)[bend right] to (-0.5,-0);
  \draw[thick, Stealth-]  (0.5,4.5)[bend left] to (0.5,-0);
\end{tikzpicture}}
\caption{Quiver for the universal flop of length $1$.}\label{Fig:UnivFlL1Quiver}
\end{figure}
Note that the final quiver, depicted in Figure~\ref{Fig:UnivFlL1Quiver}, has now two nodes less, as the U(1) of nodes 2 and 4 now disappeared from the effective theory (see \cite{Collinucci:2016hpz}). 

Let us check that the family of moduli spaces over the variety parametrized by $(\varrho_1,\varrho_2,\varrho_3)$ is the $A_3$ family we are probing. Before computing the F-terms, we integrate out the massive fields $(\phi_1-\phi_2)$ and $(\varphi_2-\varphi_4)$ obtaining:
\begin{equation}
 W^{\rm eff} = X_2\left( Y_2Z_2 - (\varphi - \varrho_2)^2 +\varrho_1\right) +   X_4\left( Y_4Z_4 - ( \varphi +\varrho_2)^2+\varrho_3 \right)  \:,
\end{equation}
with $\varphi\equiv \frac{\varphi_2+\varphi_4}{2}$. The F-terms are the following
\begin{eqnarray}
 \frac{\partial W}{\partial X_2}=0 &  : \qquad & Y_2Z_2 = (\varphi - \varrho_2)^2 -\varrho_1\:; \label{eq:FtermA3eff1}\\
 \frac{\partial W}{\partial X_4}=0 & : \qquad & Y_4Z_4 = (\varphi + \varrho_2)^2 -\varrho_3 \:; \label{eq:FtermA3eff2}\\
 \frac{\partial W}{\partial \varphi}=0 & : \qquad  & \varrho_2(X_2-X_4)-\varphi(X_2+X_4) = 0 \:;\\ 
 \frac{\partial W}{\partial Y_i}=0 & : \qquad  & X_i Z_i =0 \qquad i=2,4 \:;\\
 \frac{\partial W}{\partial Z_i}=0 & : \qquad  & X_i Y_i =0 \qquad i=2,4 \:.
\end{eqnarray}
The last equations select four branches: $\{X_2=X_4=0\}$, $\{X_2=Y_4=Z_4=0\}$, 
$\{X_4=Y_2=Z_2=0\}$ and $\{Y_2=Z_2=Y_4=Z_4=0\}$. The last three branches are incompatible with the first set of equations for generic $\varrho_1,\varrho_2,\varrho_3$. As we will see shortly, these branches are allowed on some special loci in the space $B_\varrho$, where the deformed HB develops singularities.

\subsubsection*{The $A_3$ family from the deformed HB}

Let us concentrate on the space given by $ X_2=X_4=0$ with the relations \eqref{eq:FtermA3eff1} and \eqref{eq:FtermA3eff2} and modded out by the $U(1)\times U(1)$ gauge transformations.
The gauge invariant coordinates on this moduli space are:
\begin{equation}
 U\equiv Y_2Y_4 \;, \qquad  V\equiv Z_2Z_4 \;, \qquad T_2\equiv Y_2Z_2\;, \qquad T_4\equiv Y_4Z_4\:,
\end{equation}
together with $\varphi$. These are related by $UV=T_2T_4$. The relations \eqref{eq:FtermA3eff1} and \eqref{eq:FtermA3eff2} become
\begin{equation}
T_2 =  (\varphi - \varrho_2)^2 -\varrho_1\qquad\mbox{and}\qquad  T_4 = (\varphi + \varrho_2)^2 -\varrho_3\:.
\end{equation}
Hence, the moduli space is the set parametrized by $U,V,\varphi$ modulo the relation
\begin{equation}
 U\,V =  \left( (\varphi - \varrho_2)^2 -\varrho_1 \right) \left( (\varphi + \varrho_2)^2 -\varrho_3\right) \:.
\end{equation}
This is exactly the $A_3$ family with defining equation \eqref{eq:A3family}.

\subsubsection*{Singularities of the family from the effective 3d theory}

For generic values of $\varrho_1,\varrho_2,\varrho_3$, the surface singularity is completely deformed, i.e. the HB is smooth and it is the only component of the moduli space. We expect new branches of the moduli space emerging at singularities of the HB. 
To understand which loci of $B_\varrho$ support singular surfaces, it is then enough to see when the moduli space of the probe D2-brane theory develops new branches. In this way, the effective 3d theory detects the loci where new singularities arise. 

Let us see what happens over some specific loci of $B_\varrho$:
\begin{itemize}
\item When 
$16\varrho_2^4-8\varrho_2^2(\varrho_1+\varrho_3)^2+(\varrho_1-\varrho_3)^2=0$,  the relations above allow the new branch 
\begin{equation}
Y_2=Z_2=Y_4=Z_4=0\,,\quad \varphi = \frac{\varrho_3-\varrho_1}{4\varrho_2}
\,, \quad X_4 = \left( \frac{4\varrho_2^2 + \varrho_1-\varrho_3}{4\varrho_2^2 - \varrho_1+\varrho_3} \right)X_2\:.
\end{equation}
Along this branch, the relative $U(1)$ is unbroken and $X_2$ takes any value. The monopole operators of the relative $U(1)$ can take vev subject to the relation $V_+V_-=  \left(  \frac{4\varrho_2^2 + \varrho_1-\varrho_3}{4\varrho_2^2 - \varrho_1+\varrho_3} \right)X_2^2$. I.e. we find a branch with the geometry of $\mathbb{C}^2/\mathbb{Z}_2$, that is what one expects for the CB of a D2-brane probing an $A_1$ singularity. In fact, it can be checked that on the locus under consideration, the $A_3$ surface develops an $A_1$ singularity. We have detected it by looking at the effective theory. Moreover, the $U(1)$ is untouched and one can switch on a real FI-parameter, that corresponds to a resolution of the $A_1$ singularity. This signals that this is also a singularity of the full family and that we can do a simultaneous resolution. 
\item When $\varrho_1=0$, the following new branch arises:
\begin{equation}
Y_2=Z_2=X_4=0\,\quad \varphi=\varrho_2\,,\quad Y_4Z_4=4\varrho_2^2-\varrho_3 \:,
\end{equation}
with $X_2$ unconstrained. We then see a complex 1-dimensional branch. This may look strange for a D2-brane probing a singularity. In fact, when $\varrho_1=0$ the equation of the surface manifestly develops an $A_1$ singularity at $U=V=\varphi-\varrho_2=0$. However, the background still has a T-brane deformation, that is not visible at the level of the geometry. Our effective theory is however able to detect such background, as the effective theory we obtain is the same as the one of a D2-brane probing an $A_1$ singularity with a T-brane background (see \cite{Collinucci:2016hpz,Collinucci:2017bwv}). In particular, along this branch there is no unbroken $U(1)$ (as the charged fields $Y_4,Z_4$ must take non-zero vev). Correspondingly,  one loses the possibility to deform the theory by an FI-term and there is no simultaneous resolution.
\item When $\varrho_3=0$, the following new branch arises:
\begin{equation}
Y_4=Z_4=X_2=0\,\quad \varphi=-\varrho_2\,,\quad Y_2Z_2=4\varrho_2^2-\varrho_1 \:,
\end{equation}
with $X_4$ unconstrained. All considerations done for the previous case hold here.
\item At the intersection of the first and one of the other two loci, the effective theory is the same one would obtain by probing an $A_2$ singularity with a given T-brane background. In particular there is only one $U(1)$ gauge group, with the possibility of turning on one non-zero FI-parameter: only one 2-sphere is blown up in the simultaneous resolution.
\item At the intersection $\varrho_1=\varrho_3=0$, there are new branches, but the T-brane is so severe that no $U(1)$ is left. In fact, the fiber develops an $A_1\oplus A_1$ singularity, but the sixfold is smooth.
\item At $\varrho_1=\varrho_2=\varrho_3=0$, the $A_3$ surface has the full $A_3$ singularity. Together with the HB displaying the $A_3$ singularity, there is another branch. This is what is left from the $A_3$ CB after switching on the T-brane background. There is only one $U(1)$ in the effective 3d theory with the possibility of turning on an FI-term. Again this means that the simultaneous resolution blows up just the central node.
\end{itemize}
The effective theory  tells us that the $A_3$ family has singularities only along the first locus, as it can be checked directly from its defining equation.

\section{Universal flop of length 2 from D2-branes}\label{Sec:UnFlL2}

In this section, we want to realize the universal flop of length 2 as the family of HB moduli spaces of D2-branes probing a $D_4$ singularity deformed by a specific form~of~$\Phi(\boldsymbol{\varrho})$.

\subsection{Universal flop of length 2 and the Higgs field $\Phi$}

The Higgs field that realizes the universal flop of length $\ell=2$ should allow the simultaneous resolution only of the simple root corresponding to the central node of the $D_4$ Dynkin diagram. We call this root $\alpha_4$ (see Figure~\ref{D4DynkUnivFl2}). 
This requirement implies that $\Phi$ must live in the following Levi subalgebra:
 \begin{equation}
 \mathcal{L}=A^{(\alpha_1)}_1\oplus A^{(\alpha_2)}_1\oplus A^{(\alpha_3)}_1\oplus \langle \alpha_4^*\rangle \,\subset\, D_4\:.
 \end{equation}
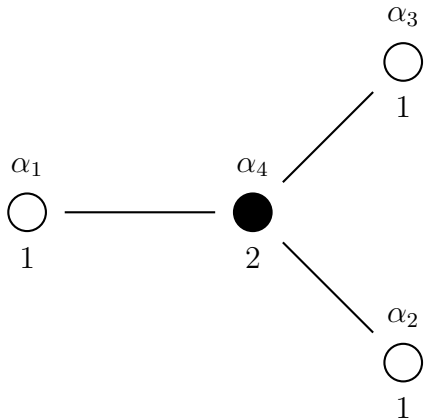
\begin{figure}[t!]
 \centering
\scalebox{1}{
\begin{tikzpicture} [place/.style={circle,draw=black!500,fill=white!20,thick,
inner sep=0pt,minimum size=5mm}, transition/.style={rectangle,draw=black!50,fill=black!20,thick,
inner sep=0pt,minimum size=10mm}, decoration={ markings,
mark=between positions 0.35 and 0.85 step 2mm with {\arrow{stealth}}}]
 \node at ( 2,-2) [place] (3) {}; \node at ( 2,-2.6) {$1$}; \node at ( 2,-1.4) {$\alpha_2$}; 
  \node at (- 3,0) [place] (1) {}; \node at (- 3,-0.6) {$1$}; \node at (-3,0.6) {$\alpha_1$}; 
 \node at (0,0) [place, fill=black] (2) {};\node at (0,-0.6) {$2$}; \node at ( 0,0.6) {$\alpha_4$}; 
  \node at (2,2) [place] (0) {};\node at (2,1.4) {$1$};\node at ( 2,2.6) {$\alpha_3$}; 
  \draw[thick] (-2.5,0)--(-0.5,0);
   \draw[thick] (0.4,0.4)--(1.6,1.6);
    \draw[thick] (0.4,-0.4)--(1.6,-1.6);
\end{tikzpicture}}
  \caption{$D_4$ Dynkin diagram. The colored node corresponds to the blown up sphere in the partial simultaneous resolution of the $D_4$ family realizing the universal flop of $\ell=2$.}\label{D4DynkUnivFl2}
\end{figure}
Following the prescription \eqref{recHiggsU} for each $A_1$ summand we have
\begin{equation}\label{PhiA1i}
\Phi|_{A_1^{(i)}} = \begin{pmatrix}
0 & 1 \\ \varrho_i & 0 
\end{pmatrix} = e_{\alpha_i}  + \varrho_i e_{-\alpha_i}  \qquad i=1,2,3\:,
\end{equation}
where $\varrho_i$ ($i=1,2,3$) is the Casimir of the $s\ell_2$ algebra $A_1^{(i)}$. Moreover $\Phi$ can have a component along $\alpha_4^*$ with coefficient $\varrho_4$.

Employing the standard basis of \cite{Collingwood}\footnote{
I.e. we choose a basis such that an element $M$ of $D_4$ is not an antisymmetric matrix but instead satisfies $M I + I M^T=0$ with
$I=\left(\begin{array}{c|c}
  \mathbb{0}   & \mathbb{1}_{4} \\
  \hline
  \mathbb{1}_{4}   & \mathbb{0} \\
\end{array}\right)$. 
} we can write $\Phi$ as the matrix
\begin{equation}\label{Flop2Phi}
\Phi=\left(
\begin{array}{cccc|cccc}
\varrho_4  & 1 & 0 & 0 & 0 & 0 & 0 & 0 \\
 \varrho_1  & \varrho_4 & 0 & 0 & 0 & 0 & 0 & 0 \\
 0 & 0 & 0 & 1 & 0 & 0 & 0 & 1 \\
 0 & 0 & \varrho_3 & 0 & 0 & 0 & -1 & 0 \\
 \hline
 0 & 0 & 0 & 0  &  -\varrho_4 & -\varrho_1  & 0 & 0 \\
 0 & 0 & 0 & 0 & -1 & -\varrho_4 & 0 & 0 \\
 0 & 0 & 0 & -\varrho_2  & 0 & 0 & 0 & -\varrho_3  \\
 0 & 0 & \varrho_2 & 0 & 0 & 0 & -1 & 0 \\
\end{array}
\right).
\end{equation}

The equation for the universal flop of $\ell=2$ can be derived by plugging \eqref{Flop2Phi} into~\eqref{Eq:DefSDn}:
\begin{equation}\label{eq:Flop2Eq}
\begin{aligned}
x^2&+y^2z  -2(\varrho_1-\varrho_4^2)(\varrho_3-\varrho_2)y \\ 
& - \frac{((z+\varrho_1-\varrho_4^2)^2+4\varrho_4^2z)((z+\varrho_3-\varrho_2)^2+4\varrho_2z)-(\varrho_1-\varrho_4^2)^2(\varrho_3-\varrho_2)^2}{z}=0
\end{aligned}
\end{equation}

For later convenience, we introduce a change of variables to realize 
the versal deformation of the $D_4$ singularity written as $X^2=Y\,Z\,(Y+Z)$. The coordinate transformation is
\begin{equation}\label{eq:Fl2ChangeCoord}
\begin{pmatrix}
x \\ y \\ z \\
\end{pmatrix}
\,\,=\,\,
\begin{pmatrix}
2X \\ Y-Z \\ Y+Z-(\varrho_1+\varrho_2+\varrho_3+\varrho_4^2) \\
\end{pmatrix}\:.
\end{equation}
In terms of the variables $X,Y,Z$, the equation for the family is
\begin{equation}\label{eq:UnivFlXYZ}
\begin{aligned}
 X^2  = &  YZ(Y+Z) - (\varrho_1+\varrho_2+\varrho_3+\varrho_4^2)YZ 
 + (\varrho_1-\varrho_2)(\varrho_3-\varrho_4^2)Y \\ 
 &+ (\varrho_1-\varrho_3)(\varrho_2-\varrho_4^2)Z 
 +(\varrho_1-\varrho_2-\varrho_3+\varrho_4^2)(\varrho_1\varrho_4^2-\varrho_2\varrho_3) \:.
\end{aligned}
\end{equation}

\subsection{D2-branes and universal flop of length $\ell=2$ }

We now describe the universal flop of length $\ell=2$ through the effective theory of a D2-brane probing a family of $D_4$ surfaces. We begin by discussing the theory at the $D_4$ singularity, followed by an analysis of how the superpotential deformation induced by $\Phi$ modifies the effective theory.

\subsubsection{D2-brane probing a $D_4$ singularity}

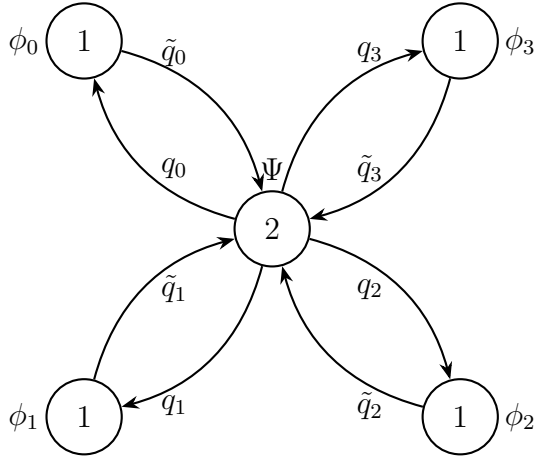
\begin{figure}[t!]
\centering
\scalebox{1}{
\begin{tikzpicture} [place/.style={circle,draw=black!500,fill=white!20,thick,
inner sep=0pt,minimum size=10mm}, transition/.style={rectangle,draw=black!50,fill=black!20,thick,
inner sep=0pt,minimum size=10mm}, decoration={ markings,
mark=between positions 0.35 and 0.85 step 2mm with {\arrow{stealth}}}]
\node at (0,0) [place] (1) {$2$};\node at (0,0.8) [] (6) {$\Psi$};
\node at ( 2.5,-2.5) [place] (2) {$1$}; \node at (3.3,-2.5) [] (8) {$\phi_2$};  \node at ( 1.3,-2.35)  {$\tilde{q}_2$}; \node at ( 1.3,-0.8)  {$q_2$}; 
 \node at ( -2.5,-2.5) [place] (3) {$1$};\node at (-3.3,-2.5) [] (9) {$\phi_1$}; \node at ( -1.3,-2.35)  {$q_1$}; \node at ( -1.3,-0.8)  {$\tilde{q}_1$}; 
  \node at ( 2.5,2.5) [place] (4) {$1$};\node at (3.3,2.5) [] (10) {$\phi_3$};  \node at ( 1.3,2.35)  {$q_3$}; \node at ( 1.3,0.8)  {$\tilde{q}_3$}; 
   \node at ( -2.5,2.5) [place] (5) {$1$};\node at (-3.3,2.5) [] (7) {$\phi_0$}; \node at ( -1.3,2.35)  {$\tilde{q}_0$}; \node at ( -1.3,0.8)  {$q_0$}; 
   \draw[->, thick, Stealth-] (1)[bend right] to (2); \draw[->, thick, -Stealth] (1)[bend left] to (2); 
  \draw[->, thick, Stealth-] (1)[bend right] to(3); \draw[->, thick, -Stealth] (1)[bend left] to(3);
  \draw[->, thick, Stealth-] (1)[bend right] to(4); \draw[->, thick, -Stealth] (1)[bend left] to(4);
  \draw[->, thick, Stealth-] (1)[bend right] to(5); \draw[->, thick, -Stealth] (1)[bend left] to(5);
\end{tikzpicture}}
\caption{$D_4$ quiver.}\label{Fig:D4quiver}
\end{figure}

The worldvolume theory of a D2-brane probing a $D_4$ singularity is a 3d $\mathcal{N}=4$ supersymmetric quiver gauge theory, with the quiver in Figure~\ref{Fig:D4quiver} and the superpotential
\begin{equation}\label{eq:D4superpot}
W= \sum_{i=0}^3\text{Tr}\left[\Psi q_i\tilde{q}_i\right]-\sum_{i=0}^3 \phi_i \tilde{q}_i q_i,
\end{equation}
where $q_i,\tilde{q}_i$ ($i=0,1,2,3$) are the bifundamental hypermultiplets, $\phi_i$ ($i=0,1,2,3$) are the scalars in the $U(1)$ vector multiplets at the external nodes and $\Psi$ is the scalar in the adjoint of $U(2)$ at the central node. In \eqref{eq:D4superpot} $q_i$ must be understood as a two-component column vector, while $\tilde{q}_i$ is a row vector.

The topological symmetry is $SO(8)$. The Cartan torus is the product of the $U(1)_T$'s of each node, whose generators can be identified with $\alpha_i^*$. The monopole operators are labeled by the charges relative to these $U(1)_T$'s, i.e. $(\mathrm{q}_0,\mathrm{q}_1,\mathrm{q}_2,\mathrm{q}_3,\mathrm{q}_4)$. Due to the decoupling of a $U(1)$ gauge multiplet, the charges of the monopoles are defined up to the shift $(\mathrm{q}_0,\mathrm{q}_1,\mathrm{q}_2,\mathrm{q}_3,\mathrm{q}_4)\mapsto (\mathrm{q}_0+1,\mathrm{q}_1+1,\mathrm{q}_2+1,\mathrm{q}_3+1,\mathrm{q}_4+2)$. To deal with this redundancy one sets $\mathrm{q}_0=0$ \cite{Bashkirov:2010kz}. The monopole operators are then written as
  $w_{\mathrm{q}_1,\mathrm{q}_2,\mathrm{q}_3,\mathrm{q}_4}$.
The topological $U(1)_T$ associated with the central node deserves further elaboration. At this node, the gauge group is $U(2)$.  To define the monopole operators, the theory is first abelianized into $U(1)\times U(1)$  \cite{Bullimore:2015lsa}. For each $U(1)$ factor, a topological $U(1)_T$ is defined: Thus, in the abelianized theory the monopole operators are labeled by two charges, that can be denoted $(\mathrm{q}_4^1,\mathrm{q}_4^2)$. However, one must quotient by the Weyl group of the original $U(2)$, which identifies the two topological $U(1)_T$. The resulting charge after the quotient is $\mathrm{q}_4\equiv \mathrm{q}_4^1+\mathrm{q}_4^2$ and the invariant monopole operator with charge $\mathrm{q}_4=1$  is given by $w_1\equiv w_{1,0}+w_{0,1}$ (where, for clarity, we have omitted the charges under nodes 1,2,3).

The moment map on the CB is given by \cite{Collinucci:2016hpz}:\footnote{We use a slightly different notation, that we justify in Appendix~\ref{AppD4mirror}.}


{\scriptsize\begin{equation*}
\bm{\mu}=\left(\begin{array}{cccc|cccc}
P_1& w_{1,0,0,0}&w_{1,0,0,1}&w_{1,0,1,1}&0&w_{1,1,1,2}&w_{1,1,1,1}&w_{1,1,0,1}\\
-w_{-1,0,0,0}&P_2&w_{0,0,0,1}&w_{0,0,1,1}&-w_{1,1,1,2}&0&w_{0,1,1,1}&w_{0,1,0,1}\\
-w_{-1,0,0,-1}&-w_{0,0,0,-1}&P_3&w_{0,0,1,0}&-w_{1,1,1,1}&-w_{0,1,1,1}&0&w_{0,1,0,0}\\
-w_{-1,0,-1,-1}&-w_{0,0,-1,-1}&-w_{0,0,-1,0}&P_4&-w_{1,1,0,1}&-w_{0,1,0,1}&-w_{0,1,0,0}&0\\\hline
0&w_{-1,-1,-1,-2}&w_{-1,-1,-1,-1}&w_{-1,-1,0,-1}&-P_1&w_{-1,0,0,0}&w_{-1,0,0,-1}&w_{-1,0,-1,-1}\\
-w_{-1,-1,-1,-2}&0&w_{0,-1,-1,-1}&w_{0,-1,0,-1}&-w_{1,0,0,0}&-P_2&w_{0,0,0,-1}&w_{0,0,-1,-1}\\
-w_{-1,-1,-1,-1}&-w_{0,-1,-1,-1}&0&w_{0,-1,0,0}&-w_{1,0,0,1}&-w_{0,0,0,1}&-P_3&w_{0,0,-1,0}\\
-w_{-1,-1,0,-1}&-w_{0,-1,0-1}&-w_{0,-1,0,0}&0&-w_{1,0,1,1}&-w_{0,0,1,1}&-w_{0,0,1,0}&-P_4\end{array}\right)
\end{equation*}}
with
\begin{equation}\label{eq:P1P2P3P4}
(P_1,P_2,P_3,P_4)=( \phi_0-\phi_1,-2\psi+\phi_0+\phi_1,2\psi-\phi_2-\phi_3,\phi_3-\phi_2) \:.
\end{equation}

\subsubsection{Monopole deformations and the effective theory}

The field $\Phi$ in \eqref{Flop2Phi} generates the following superpotential deformation\footnote{The $1/2$ normalization factor  is due to the matrix representation we are using for $\Phi$ and $\mu$.} $\delta W=\frac12 \tr[\Phi\bm{\mu}]$ on the D2-brane probe worldvolume:
\begin{equation}
\delta W = -w_{-1,0,0,0} -w_{0,-1,0,0}-w_{0,0,-1,0}   +\varrho_1w_{1,0,0,0} +\varrho_2 w_{0,1,0,0} + \varrho_3w_{0,0,1,0} +2\varrho_4 (\phi_0-\psi)\:. 
\end{equation}
We note that the expression is  symmetric in the exchange of the three external simple roots $\alpha_1,\alpha_2,\alpha_3$. 

\begin{figure}[t!]
\centering
\scalebox{1}{
\begin{tikzpicture} [place/.style={circle,draw=black!500,fill=white!20,thick,
inner sep=0pt,minimum size=10mm}, transition/.style={rectangle,draw=black!50,fill=black!20,thick,
inner sep=0pt,minimum size=10mm}, edge/.style={stealth}]
   \node at (8,0) [place] (1) {$2$}; \node at (8,0.75) [] (4) {$\psi$};
     \node at ( 4,2.7) [place] (2) {$1$};\node at (4,3.45) [] (5) {$\phi_0$};
     \node at (9.8,-0.3) [] (6) {$\mathfrak{M}_3$};  \node at (8.5,-1.7) [] (7) {$\mathfrak{M}_2$}; 
      \node at (6.4,-0.7) [] (8) {$\mathfrak{M}_1$};   \node at (6,2.6) {$\tilde{q}_0$}; \node at (6,0.65) {$q_0$};
     \draw[thick] (1) [in=-35,out=20,loop, -Stealth] to (1); 
     \draw[thick] (1) [in=-105,out=-45,loop, -Stealth] to (1); 
     \draw[thick] (1) [in=-175,out=-115,loop, -Stealth] to (1); 
     \draw[thick] (1) [in=35,out=105,loop, -Stealth] to (1); \node at (8.5, 1.75) {$\tilde{\Psi}$};
  \draw[thick, Stealth-]  (1)[bend right] to (2);
   \draw[thick, -Stealth]  (1)[bend left] to (2);
\end{tikzpicture}}
\caption{Quiver for the effective theory.}\label{Fig:EffThFlopL2quiver}
\end{figure}
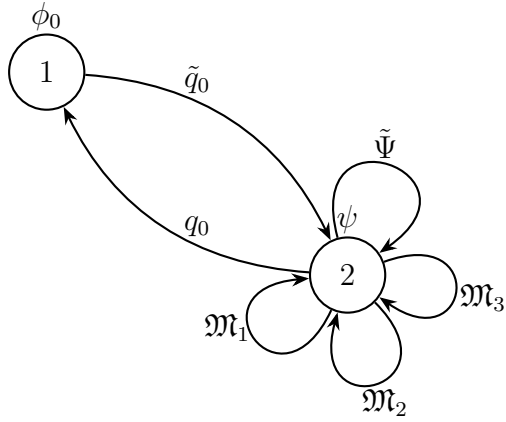

We now apply the algorithm explained in Section~\ref{Sec:A3example}:
\begin{enumerate}
\item Let us consider node 1. Nodes 2 and 3 behave in the same way due to the symmetry between the three external nodes. The isolated theory is $U(1)$ with two flavors and superpotential
\begin{equation}
W_1 = - \phi_1 \tr \mathcal{M}_1 -w_{-1,0,0,0}+\varrho_1 w_{1,0,0,0} \:.
\end{equation}
This is of the form \eqref{eq:WBA1case} encountered in the $A_1$ case.
\item Following  the same steps as in the previous examples, we obtain for each node a modified XYZ theory with superpotential
\begin{equation}
W_i^{\rm eff} = X_i \det \mathfrak{M}_i + \varrho_i \, X_i \qquad\mbox{with}\qquad 
\mathfrak{M}_i \equiv \begin{pmatrix}
 \varphi_i & -Y_i \\ Z_i & -\varphi_i  \\
\end{pmatrix} \qquad i=1,2,3\:
\end{equation}
and flavor symmetry $SU(2)$. 
\item We now couple these isolated theories to the nearby nodes. For each node, we need to add 
\begin{equation}
 \tr \left[ \Psi \mathfrak{M}_i \right]  = \tr \left[ \tilde{\Psi} \mathfrak{M}_i \right]   \qquad\mbox{with}\quad     \tilde{\Psi}\equiv \Psi-\psi\mathbf{1} \,\,\mbox{ and }\,\, \psi\equiv \frac{1}{2}\text{Tr}(\Psi)\:,
\end{equation}
and where on the right-hand side only the traceless part of $\Psi$ survives, as $\mathfrak{M}_i$ is traceless.
\end{enumerate}
Adding all the terms together, we obtain the effective superpotential
\begin{equation}
W_{\rm eff} = \sum_{i=1}^3 X_i\left(   \det \mathfrak{M}_i + \varrho_i   \right) 
 + \tr \left[ \tilde{\Psi} \left( q_0\tilde{q}_0 + \sum_{i=1}^3  \mathfrak{M}_i \right)  \right]   + \hat\psi \left(\tilde{q}_0q_0 - 2\varrho_4 \right) \:,
\end{equation}
with $\hat{\psi}\equiv \psi-\phi_0$. The quiver of this effective theory is given in Figure~\ref{Fig:EffThFlopL2quiver}.

The F-terms are the following:
\begin{eqnarray}
 \frac{\partial W}{\partial X_i}=0 &  : \qquad &  \det \mathfrak{M}_i = - \varrho_i  \qquad i=1,2,3 \label{eq:FtermD4fam1}\\
 \frac{\partial W}{\partial \hat{\psi}}=0 &  : \qquad & \tilde{q}_0 q_0 =2 \varrho_4 \label{eq:FtermD4fam2}  \\
 \frac{\partial W}{\partial \tilde{\Psi}}=0 &  : \qquad & \sum_{i=1}^3  \mathfrak{M}_i + \left(q_0\tilde{q}_0 - \frac12 (\tilde{q}_0 q_0) \mathbb{1} \right)  = 0 \label{eq:FtermD4fam3}\\
 \frac{\partial W}{\partial \mathfrak{M}_i}=0 &  : \qquad & \tilde{\Psi} - X_i\mathfrak{M}_i=0 \qquad i=1,2,3 \label{eq:FtermD4fam4} \\ 
 \frac{\partial W}{\partial q_0}=0 &  : \qquad & \tilde{q}_0\left(\tilde{\Psi}+\hat{\psi}\mathbb{1}\right) = 0  \label{eq:FtermD4fam5} \\
 \frac{\partial W}{\partial \tilde{q}_0}=0 &  : \qquad & \left(\tilde{\Psi}+\hat{\psi}\mathbb{1}\right) q_0 = 0 \:.\label{eq:FtermD4fam6}
\end{eqnarray}

The 3d gauge theory with quiver in Figure~\ref{Fig:EffThFlopL2quiver} has also a D-term potential. 
The gauge group is $U(1)\times U(2)$.
The D-term relations are:
\begin{equation}\label{eq:DtermsU1U2th}
\begin{aligned}
&q_0^\dagger q_0-\tilde{q}_0 \tilde{q}_0^\dagger = 2\xi \:,\\
&q_0 q_0^\dagger-\tilde{q}_0^\dagger \tilde{q}_0  + [\tilde{\Psi},\tilde{\Psi}^\dagger] + \sum_{i=1}^3[\mathfrak{M}_i,\mathfrak{M}_i^\dagger]= \xi \mathbb{1}_2 \:.\\
\end{aligned}
\end{equation}
where $\xi$ is the real FI-parameter associated with the $U(1)$ generator $T_1-T_2$, where $T_i$ generates the diagonal $U(1)$ factor of the gauge group at the $i$-th node in Figure~\ref{Fig:EffThFlopL2quiver} (the diagonal $U(1)$ generated by $T_1+T_2$ decouples as no field is charged under it).

\

As we will see shortly, these relations can potentially give rise to distinct branches of the moduli space. For generic values of $\boldsymbol{\varrho}=(\varrho_1, \dots, \varrho_4)$, the moduli space of the effective 3d theory has a single branch, corresponding to the deformed HB. 
The family of these spaces is the universal flop of length $\ell=2$. 
At special points in the space $B_\varrho$, parametrized by $\varrho_1, \dots, \varrho_4$, the fiber surface develops singularities, leading to new branches in the moduli space. By analyzing the $U(1)$ factors in the gauge group of the effective theory, one can determine whether the universal flop sixfold is singular at these points in $B_\varrho$. We will examine this in detail in the following.

\subsubsection{The universal flop of length 2 as a family of deformed HB's}\label{Sec:EqFamilyHBUnFl2}

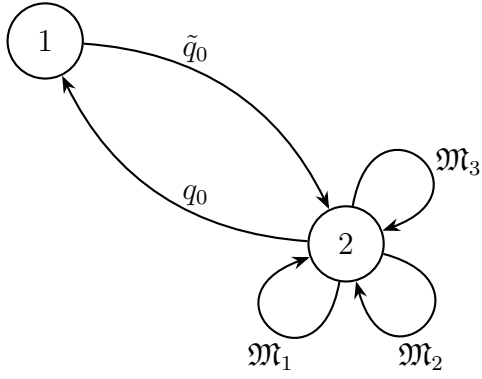
\begin{figure}[t!]
\centering
\begin{tikzpicture} [place/.style={circle,draw=black!500,fill=white!20,thick,
inner sep=0pt,minimum size=10mm}, transition/.style={rectangle,draw=black!50,fill=black!20,thick,
inner sep=0pt,minimum size=10mm}, edge/.style={stealth}]
   \node at (8,0) [place] (1) {$2$}; \node at (7.97,0.7) [] (4) {$$};
     \node at ( 4,2.7) [place] (2) {$1$};\node at (5.5,2.75) [] (5) {$$};
     \node at (9.5,1.1) [] (6) {$\mathfrak{M}_3$};  \node at (9,-1.5) [] (7) {$\mathfrak{M}_2$}; 
      \node at (7,-1.5) [] (8) {$\mathfrak{M}_1$};   \node at (6,2.6) {$\tilde{q}_0$}; \node at (6,0.65) {$q_0$};
     \draw[thick] (1) [in=20,out=80,loop, -Stealth] to (1); 
     \draw[thick] (1) [in=-75,out=-15,loop, -Stealth] to (1); 
     \draw[thick] (1) [in=-160,out=-100,loop, -Stealth] to (1); 
  \draw[thick, Stealth-]  (1)[bend right] to (2);
   \draw[thick, -Stealth]  (1)[bend left] to (2);
\end{tikzpicture}
\caption{Quiver for the universal flop of length 2.}\label{Fig:UnFl2quiver}
\end{figure}

Let us consider the case in which $\varrho_1,...,\varrho_4$ take generic non-zero values. As we will prove next, in this case the relations \eqref{eq:FtermD4fam1}-\eqref{eq:FtermD4fam6}
imply that $X_1=X_2=X_3=0$,  $\tilde{\Psi}=0$ and $\hat{\psi}=0$.
We are therefore left with a moduli space parametrized by the traceless matrices $\mathfrak{M}_i$ ($i=1,2,3$) and the bifundamentals $q_0$ and $\tilde{q}_0$ organized in an effective quiver, that we show in Figure~\ref{Fig:UnFl2quiver}. The maps in the quiver are subject to the relations
\begin{equation}
  \sum_{i=1}^3  \mathfrak{M}_i + q_0\tilde{q}_0 = \varrho_4 \mathbb{1} \qquad\mbox{and}\qquad  \det \mathfrak{M}_i = - \varrho_i  \qquad i=1,2,3    \:.
\end{equation}
These relations, together with the quiver in Figure~\ref{Fig:UnFl2quiver}, reproduce the `universal flopping algebra of length $\ell=2$' derived by \cite{Karmazyn:2017aa},
from which the universal flop of $\ell=2$ can be recovered by a moduli construction.\footnote{The precise match with Example 4.23 in \cite{Karmazyn:2017aa} works in the following way: 1) the quiver is the same; 2) the relations in \cite{Karmazyn:2017aa} imply that the self-arrows of the rank-2 node are traceless; 3) one uses that for traceless $2\times 2$ matrices $\tilde{M}$, the relation $\tilde{M}^2=-$det$(\tilde{M})\,\mathbb{1}_2$ holds.}

The invariants that parametrize this branch of the moduli space consist of single trace chiral operators, which correspond to closed loops in the quiver. A convenient choice for a basis is
\begin{equation}
A_i\equiv \tr\left[\mathfrak{M}_0\mathfrak{M}_i\right], \qquad B\equiv \tr\left[\mathfrak{M}_0\mathfrak{M}_2\mathfrak{M}_3\right]\:,
\end{equation}
where we defined $\mathfrak{M}_0\equiv q_0\tilde{q}_0$.

Crucially $A_1,\,A_2,\,A_3$ and $B$ are not independent. In particular, \eqref{eq:FtermD4fam3} implies that: 
\begin{equation}
\sum_i A_i = -2\varrho_4^2 \:.
\end{equation}
Moreover, there is an additional relation involving $B$, which we derive in Appendix~\ref{App:UnFl2relquiver}:
\begin{equation}\label{eq:BA1A2A3equation}
\begin{aligned}
A_1A_2A_3 =& -B^2 +2B\varrho_4(A_3+\rho_4^2+\rho_1-\rho_2-\rho_3) \\
&-(\rho_2A_3+\rho_3A_2)(A_1+2\rho_4^2)-A_2A_3(\rho_4^2+\rho_1)-4\rho_4^2\rho_2\rho_3\:.
\end{aligned}
\end{equation}
We write $A_1$ in terms of $A_2$ and $A_3$ by using the first relation, and make the following redefinition of the gauge invariant coordinates:
\begin{eqnarray}
A_2 &=& Y-\varrho_2-\varrho_4^2 \;, \\
A_3 &=& Z-\varrho_3-\varrho_4^2 \;, \\
B &=& X-(Y+Z-\varrho_1)\varrho_4+\varrho_4^3 \;. 
\end{eqnarray}
with $A_1=-Y-Z+\varrho_2+\varrho_3$. 
Plugging these redefinitions into \eqref{eq:BA1A2A3equation}, we obtain a space parametrized by $X,Y,Z$ subject to the relation
\begin{equation}\label{eq:UnivFlFromQuiverXYZ}
\begin{aligned}
 X^2  = &  YZ(Y+Z) - (\varrho_1+\varrho_2+\varrho_3+\varrho_4^2)YZ 
 + (\varrho_1-\varrho_2)(\varrho_3-\varrho_4^2)Y \\ 
 &+ (\varrho_1-\varrho_3)(\varrho_2-\varrho_4^2)Z 
 +(\varrho_1-\varrho_2-\varrho_3+\varrho_4^2)(\varrho_1\varrho_4^2-\varrho_2\varrho_3) \:.
\end{aligned}
\end{equation}
This is exactly the equation \eqref{eq:UnivFlXYZ}, that defines the {\it universal flop of length $\ell=2$} as a hypersurface in the seven dimensional ambient space $\mathbb{C}^3_{XYZ}\times B_\varrho$.

\subsubsection{Singularities of the sixfold from the effective 3d theory}

We now search for loci in $B_\varrho$ where the moduli space of the effective 3d theory develops new branches. As explained in Section~\ref{Sec:GenericADEalgSuperpotDef}, at these loci, the ADE surface develops a singularity. 
From the quiver in Figure~\ref{Fig:EffThFlopL2quiver}, we observe that there is only one relevant $U(1)$ subgroup, which is the relative $U(1)$ between the $U(1)$ factors at each node  (the diagonal $U(1)$ is decoupled). Hence the theory has only one real FI-parameter that can be activated (leading to the simultaneous resolution of the family). 
To determine whether the loci in $B_\varrho$ that support new branches of the moduli space correspond to singularities in the sixfold family, it is crucial to verify if a $U(1)$ gauge group is preserved at the intersection of these branches.

There are two loci of $B_\varrho$ where we detect a singularity of the fiber that is blown up in the simultaneous resolution of the family:
\begin{itemize}
\item When $\varrho_4=0$, the relations \eqref{eq:FtermD4fam1}-\eqref{eq:FtermD4fam6} allow to set $q_0=\tilde{q}_0=0$; the corresponding new branch is determined by the following relations: 
\begin{equation}\label{eq:Ftermrho4zeroD4fam} 
   \sum_{i=1}^3  \mathfrak{M}_i  = 0\,,\qquad  \det \mathfrak{M}_i = - \varrho_i \,\,\mbox{ and }\,\, 
 \tilde{\Psi} - X_i\mathfrak{M}_i=0 \quad (i=1,2,3) \:.
\end{equation}
For generic $\varrho_1,\varrho_2,\varrho_3$, \eqref{eq:Ftermrho4zeroD4fam} imply $X_i=0$ and $\tilde{\Psi}=0$.\footnote{The matrices $\mathfrak{M}_i$ are invertible. $\tilde{\Psi}=X_i\mathfrak{M}_i$ says that the $X_i$ are either all zero or all non-zero. In the second case, the $\mathfrak{M}_i$ should be proportional to each other. But this is in conflict with $\sum_{i=1}^3  \mathfrak{M}_i  = 0$ for generic 
$\varrho_1,\varrho_2,\varrho_3$.}
The first two relations in  \eqref{eq:Ftermrho4zeroD4fam} imply that $\mathfrak{M}_i$ are fixed (up to gauge transformations) to a specific value determined by the parameters $\varrho_1,\varrho_2,\varrho_3$.\footnote{The first relation implies $\mathfrak{M}_1=-\mathfrak{M}_2-\mathfrak{M}_3$; the other three relations determine all the gauge invariants of two traceless $2\times 2$ matrices. In particular, $\det\mathfrak{M}_2=-\varrho_2$, $\det\mathfrak{M}_3=-\varrho_3$ and $\tr \mathfrak{M}_2\mathfrak{M}_3=- \det(\mathfrak{M}_2+\mathfrak{M}_3)+\det\mathfrak{M}_2+\det\mathfrak{M}_3=\varrho_1-\varrho_2-\varrho_3$,
where we used the fact that for traceless matrices $\mathfrak{M}_i$ the following relation holds: $\det\mathfrak{M}_i=-\frac12 \tr( \mathfrak{M}_i^2)$.}
The only field that can vary along the new branch is $\hat{\psi}$. 
Along this branch the relative $U(1)$ of the two nodes is unbroken. Its monopole operators $V_\pm$ can get non zero vev as well, subject to the relation $V_+V_-=\hat{\psi}^2$. The new branch has then the geometry of a $\mathbb{C}^2/\mathbb{Z}_2$, like the CB of a $U(1)$ theory with two flavors. 

In fact, at the intersection of the new branch with the HB, the effective theory {\it is} a 3d $\mathcal{N}=4$ supersymmetric $U(1)$ gauge theory with two flavors: integrating out the massive fields $\tilde{\Psi},\mathfrak{M}_i,X_i$ one obtains the superpotential 
$W_{\rm eff}'=\hat\psi \, \tilde{q}_0q_0$.
This is compatible with the fact that the D2-brane is probing an $A_1$ singularity. In fact, when $\varrho_4=0$ the equation \eqref{eq:UnivFlFromQuiverXYZ} takes the form
\begin{equation}
X^2=\varrho_3(Y-\varrho_2)^2+\varrho_2(Z-\varrho_3)^2+(Y-\varrho_2)(Z-\varrho_3)(Y+Z-\varrho_1)\;,
\end{equation}
that has a manifest $A_1$ singularity at $X=Y-\varrho_2=Z-\varrho_3=0$. 

The D-term condition of the effective theory can be derived by looking at \eqref{eq:DtermsU1U2th} and setting the massive fields as above, i.e. $\tilde{\Psi}=0$ and $\mathfrak{M}_i$ traceless invertible matrices, that are then diagonalizable by gauge transformations. This leaves the following D-term:
\begin{equation}
q_0^\dagger q_0-\tilde{q}_0 \tilde{q}_0^\dagger = 2\xi \:.
\end{equation}
Switching on a non-zero real FI-parameter $\xi$ blows up the $A_1$ singularity of the HB. This is the FI-parameter of the theory with quiver in Figure~\ref{Fig:EffThFlopL2quiver}, hence the simultaneous resolution of the family sixfold blows up the $A_1$ simple root.

The effective theory has detected a singular locus in the sixfold, where a $\mathbb{CP}^1$ is blown up in the simultaneous resolution of the family.

\item When 
\begin{equation}\label{eq:UnFl2SecondBranch}
\begin{aligned}
64\varrho_4^2\varrho_1\varrho_2\varrho_3=&\left[ \varrho_1^2+\varrho_2^2+\varrho_2^2-2\varrho_1\varrho_2-2\varrho_1\varrho_3-2\varrho_2\varrho_3 \right.\\
&\left.-2\varrho_4^2(\varrho_1+\varrho_2+\varrho_3)+\varrho_4^4
\right]^2 \:,
\end{aligned}
\end{equation}
a new branch emerges, as we now show. 
To present in a cleaner shape what happens at one point along this locus in $B_\varrho$, let us write
\begin{equation}\label{eq:rhoVSr}
 \varrho_1=r_1^2\,,\qquad \varrho_2=r_2^2\,,\qquad \varrho_3=r_3^2\,,\qquad \varrho_4=r_4\,.
\end{equation}
We stress that the holomorphic coordinates on $B_\varrho$ are still $\varrho_1,...,\varrho_4$.
The locus \eqref{eq:UnFl2SecondBranch} can then be written as points where
\begin{equation}\label{eq:secondBranchWithri}
\pm r_1\pm r_2 \pm r_3 + r_4 = 0 \:.
\end{equation}
The different choices of signs select distinct regions of the locus \eqref{eq:UnFl2SecondBranch}(that is connected).

At the locus \eqref{eq:UnFl2SecondBranch}, the relations \eqref{eq:FtermD4fam1}-\eqref{eq:FtermD4fam6} allow the $X_i$ to be non-zero and consequently among the other fields $\hat{\psi}\neq 0$ is permitted. 
When the $X_i$ are non-zero (they either all vanish or none do), $\mathfrak{M}_i$ ($i=1,2,3$) and $\tilde{\Psi}$ are all proportional to each other. Hence, we can partially fix the gauge by diagonalizing them simultaneously.\footnote{They are invertible and traceless, meaning they have distinct non-zero eigenvalues.} The matrices $\mathfrak{M}_i$ are then determined up to a sign by \eqref{eq:FtermD4fam1}, and $\tilde{\Psi}$ depends on one complex scalar $\psi_3$:
\begin{equation}\label{eq:gaugeFixSecondLocUnFl}
  \mathfrak{M}_i = \sigma_i \, \begin{pmatrix}
  r_i & 0 \\ 0 & -r_i
  \end{pmatrix} \quad\mbox{with}\quad \sigma_i=\pm \,, \qquad\mbox{and}\qquad
  \tilde{\Psi} = \begin{pmatrix}
  \psi_3 & 0 \\ 0 & -\psi_3
  \end{pmatrix}\:.
\end{equation}
Due to \eqref{eq:FtermD4fam4}, we immediately see that $X_i=\frac{\sigma_i}{r_i}\psi_3$. Moreover, \eqref{eq:FtermD4fam6} says that $q_0$ is an eigenvector of $\tilde{\Psi}$ with eigenvalue $\hat{\psi}$. After the gauge fixing \eqref{eq:gaugeFixSecondLocUnFl}, this implies two possibilities for $q_0$ and $\hat{\psi}$:
\begin{equation}
  q_0=\begin{pmatrix}
   q_0^1 \\ 0 \\
  \end{pmatrix} \,\mbox{ and } \, \hat{\psi}=-\psi_3 \qquad \mbox{or}\qquad
  q_0=\begin{pmatrix}
   0 \\ q_0^2 \\
  \end{pmatrix} \,\mbox{ and } \, \hat{\psi}=\psi_3\:.
\end{equation}
Analogous considerations can be taken from the relation \eqref{eq:FtermD4fam5}:
\begin{equation}
  \tilde{q}_0=\begin{pmatrix}
   \tilde{q}_0^1 & 0 \\
  \end{pmatrix} \,\mbox{ and } \, \hat{\psi}=-\psi_3 \qquad \mbox{or}\qquad
  \tilde{q}_0=\begin{pmatrix}
   0 & \tilde{q}_0^2 \\
  \end{pmatrix} \,\mbox{ and } \, \hat{\psi}=\psi_3\:.
\end{equation}
Let us consider the case $\hat{\psi}=-\psi_3$ (the second case is obtained from the first one by applying the Weyl group of the $SU(2)$ of the second node). The relation \eqref{eq:FtermD4fam2} says $\tilde{q}_0^1q_0^1=2r_4$.

We can now plug everything into \eqref{eq:FtermD4fam3}, obtaining
$
\sum_{i=1}^3 \sigma_ir_i  + r_4 = 0 
$, 
i.e. all the relations are compatible with each other only over the locus \eqref{eq:UnFl2SecondBranch}. Moreover, the choice of signs $\sigma_i$ determines which region of the locus we are on.

Summing up, over the locus \eqref{eq:UnFl2SecondBranch} in $B_\varrho$, the effective 3d theory moduli space has a new branch parametrized by $\psi_3$. The vev of the fields break the $U(1)\times U(2)$ gauge group to $U(1)\times U(1)$, with one combination of them that decouples. 

After performing a series of involved steps, which we omit here for brevity, one can verify that, by integrating out the fields that become massive for each value of the moduli, the resulting effective theory at the intersection of the new branch with the HB is a $U(1)$ gauge theory with two flavors (one from $q_0, \tilde{q}_0$ and the other from the off-diagonal elements of a combination of $\mathfrak{M}_i$).
This is compatible with the fact that over the locus \eqref{eq:UnFl2SecondBranch} the surface fiber develops an $A_1$ singularity. For example, plugging   \eqref{eq:rhoVSr} with $r_4=r_1+r_2+r_3$  into \eqref{eq:UnivFlFromQuiverXYZ} and shifting the coordinates as  $Y=\mathrm{y}+(r_1+r_3)^2$ and $Z=\mathrm{z}+(r_1+r_2)^2$, one obtains
\begin{equation}
X^2= (r_1+r_2)^2 \mathrm{y}^2+(r_1+r_3)^2 \mathrm{z}^2+ 2(y+z+r_1^2+r_1(r_2+r_3)-2r_2r_3)\mathrm{y}\,\mathrm{z}\:,
\end{equation}
that manifestly has an $A_1$ singularity at $X=\mathrm{y}=\mathrm{z}=0$.

Also in this case,  switching on the FI-term in the original theory with quiver in Figure~\ref{Fig:EffThFlopL2quiver} will produce a real FI-term in the effective theory just described, that will blow up the $A_1$ singularity of the HB. 

The effective theory has detected in this way a singularity of the sixfold along the locus \eqref{eq:UnFl2SecondBranch}.

\end{itemize}

When $\varrho_i = 0$ for one of $i \in {1, 2, 3}$, the effective theory develops a new branch beyond the HB. Let us take $i=1$ as a reference (the same arguments apply for $i=2, 3$).  In this case, $\mathfrak{M}_1$ can acquire a zero vev, allowing $X_1$ to take any value. The new branch is parametrized by this field. The non-zero vevs for the fields $\mathfrak{M}_{2,3}$ and $q_0,\tilde{q}_0$, that are forced by the relations \eqref{eq:FtermD4fam1}-\eqref{eq:FtermD4fam6}, break the gauge group completely at any point in the moduli space, leaving no $U(1)$ factor. This indicates that, along this locus, the sixfold family develops no singularities (even though the fiber surface exhibits an $A_1$ singularity, as can be confirmed from the equation).

There are also relevant subloci: 
\begin{itemize}
\item The two singular loci described above intersect at
\begin{equation}
\varrho_4=0  \qquad\mbox{and}\qquad \varrho_1^2+\varrho_2^2+\varrho_2^2-2\varrho_1\varrho_2-2\varrho_1\varrho_3-2\varrho_2\varrho_3=0\:.
\end{equation}
Studying the moduli space of the effective theory over this locus in parameter space, one discovers that the ADE surface has an $A_2$ singularity. In fact, switching on the only possible real FI-parameter one blows up two $\mathbb{CP}^1$'s \cite{Collinucci:2019fnh}.
\item The origin of the parameter space is at $\varrho_1 = \varrho_2 = \varrho_3 = \varrho_4 = 0$. At this point, the effective theory corresponds to a D2-brane probing a $D_4$ singularity with a T-brane background. The blown-up $\mathbb{CP}^1$ corresponds to the central node of the $D_4$ diagram and has length 2.
\end{itemize}

This reveals the structure of the simultaneous resolution of the universal flop of length 2 (here derived by analyzing the moduli space of an effective 3d theory of a D2-brane probing deformed $D_4$ surfaces). There are two codimension-1 loci in $B_\varrho$ where a single $\mathbb{CP}^1$ is blown up. At the intersection of these loci, the exceptional locus consists of two $\mathbb{CP}^1$'s that intersect at a single point. At the origin of $B_\varrho$, the two spheres coincide, forming a degree-two $\mathbb{CP}^1$ (see e.g. \cite{Collinucci:2019fnh}). 

\subsection{Monopole operators as D2 states}

The D2-branes wrapped on vanishing spheres give rise to particles propagating in six dimensions. The D2-branes are charged under the Ramond-Ramond $C_3$ 3-form potential. In an ADE surface there are $r$ 2-forms $\omega_i$ $i=1,...,r$ such that their integral over a basis of simple roots is
\begin{equation}\label{eq:omegaialphaj}
 \int_{\alpha_j}\omega_i = \delta_i^j \:.
\end{equation}
$C_3$ can be expanded as $C_3\sim \sum_{i=1}^r A^i \wedge \omega_i$, with $A_i$ a gauge field propagating in six dimensions. These are background fields from the D2-brane 3d theory point of view. Given a D2-brane wrapping a root $\alpha$, its charge under the $U(1)$ gauge group generated by $\omega_i$ is 
\begin{equation}
\mathrm{q}_i(D2_\alpha) = \int_\alpha \omega_i \:. 
\end{equation}

The simple roots are seen as linear functional acting on the space of two-forms, that is then identified with the Cartan subalgebra of the corresponding ADE Lie algebra. The relations \eqref{eq:omegaialphaj} then mean that $\omega_i=\alpha^*_i$, i.e. the dual basis of the simple roots. 

In \cite{Collinucci:2016hpz}, the authors 
studied the 3d theory living on the worldvolume of D2-branes probing the ADE singularity and extending in the non-compact $\mathbb{R}^{1,2}$ spacetime. As explained before, the symmetry associated with the $A^i$'s is the topological symmetry in the 3d theory.
In this context \cite{Collinucci:2016hpz} 
showed that monopole operators map to states of D2-branes wrapping vanishing spheres: the existence of a monopole operator with charge $\mathrm{q}$ with respect to one $\alpha^*_j$ means that the string theory produces a D2-state with charge $\mathrm{q}$ under the $U(1)$ symmetry with  gauge field $A^j$.
In the 3d theory supported on the probe D2-brane (that should be distinguished from the D2-branes generating charged states), the $A^i$'s are background gauge fields for the topological symmetry and the D2-brane wrapping vanishing spheres are mapped to monopole operators with the proper charges under the Cartan torus of the topological ADE symmetry.

Now, let us come to the universal flop of length two. 
This has been obtained by switching on a particular monopole deformation on the worldvolume theory of a D2-brane probing a $D_4$ singularity. Before turning on the deformation, the topological symmetry was $SO(8)$. The new terms in the superpotential break explicitly this symmetry to a $U(1)_T$ symmetry generated by $\alpha_4^*$. This is the topological symmetry that can be read off from the quiver of the effective theory in Figure~\ref{Fig:EffThFlopL2quiver}. 
The abelianization of the gauge group has gauge group $U(1)\times U(1)^2$, with one combination of them that decouples. The topological charges of the abelianized theories are $(\mathrm{q}_0,\mathrm{q}_4^1,\mathrm{q}_4^2)$, that are defined up to a shift, due to the decoupled $U(1)$. To fix the ambiguity, we set $\mathrm{q}_0=0$. The charge under the topological $U(1)$ is $\mathrm{q}_4\equiv\mathrm{q}_4^1+\mathrm{q}_4^2$.

At a generic point of $B_\varrho$, the effective 3d theory lacks a $U(1)$ gauge group, resulting in the absence of monopole operators. Consequently, we conclude that there are no charged states coming from D2-branes wrapping vanishing cycles (the surface is smooth here).
On the locus $\varrho_4=0$ there is a preserved $U(1)$; in the abelianized theory its associated monopole operators have charges $(\mathrm{q}_0,\mathrm{q}_4^2,\mathrm{q}_4^2)=(\pm 1,0,0)\cong \pm (0,1,1)$, i.e. they are the monopole operators $w_{\pm 2}$. On the second locus \eqref{eq:UnFl2SecondBranch}, the monopole operators associated with the surviving $U(1)$ generator have charges $(\mathrm{q}_0,\mathrm{q}_4^2,\mathrm{q}_4^2)=(0,\pm 1,0)$ (one needs to consider the Weyl transformed as well); they are the monopole operators $w_{\pm 1}$. We conclude that on the first locus we have a D2-state of charge $2$, while on the second locus we have a D2-state of charge $1$.
At the origin of $B_\varrho$, we have both types of states.

Following~\cite{Collinucci:2019fnh}, when we let all $\varrho_i$  depend on a complex coordinate, this generates  a three-fold known as the Morrison-Park threefold \cite{Morrison:2012ei}. Our analysis above reproduces the structure of charge one and charge two states coming from reducing type IIA/M-theory on such a threefold (see \cite{Collinucci:2019fnh} for a review). The same can be done for fourfolds, constructed by making $\varrho_i$ depend on two complex variables. We leave for future work the analysis of the fourfolds.

This analysis can also be applied to CY threefold flops of $\ell=2$. The structure of charged states is more intricate than for the Morrison-Park threefold and it has been preliminarily analyzed for the Laufer threefold in \cite{Collinucci:2018aho}, whose results match with the types of monopole operators at the origin of the $D_4$-family that realize the universal flop of $\ell=2$.

\section{Conclusions}\label{Sec:Conclusions}

In this paper, we have explored a physics-driven perspective on \textit{universal flops}, a special class of algebraic varieties. As we explained before, these are families of deformed ADE surfaces with partial simultaneous resolution, which have allowed for an exhaustive classification of CY threefolds with simple flops. The ADE families define fibrations over the deformation parameter space $B_{\rho}$. At the origin of $B_{\rho}$, the fiber displays the full ADE singularity, but the resolution of the total space of the family blows up only one $\mathbb{CP}^1$. Outside the origin, the fibers are smooth except for special regions of the parameter space. For specific values of the parameters, the fiber may still be singular. However, its resolution may or may not be obstructed. 

We have shown that this structure can be fully reconstructed and analyzed by studying the moduli space of a D2-brane probing these geometric backgrounds. In Type IIA string theory, 
the deformation can be encoded into an adjoint `Higgs field' background $\Phi(\boldsymbol{\varrho})$, that encodes all the geometric features of the ADE family \cite{Collinucci:2021ofd}. On the D2-brane worldvolume, this scalar $\Phi(\boldsymbol{\varrho})$ couples linearly to the moment map of the ADE Lie algebra, introducing monopole operator deformations and complex FI terms. These deformations play a crucial role in lifting the CB and deforming the HB. As a result, the ADE family emerges as a family of deformed Higgs branches,  parametrized by the  deformation parameters $\boldsymbol{\varrho}$. For special values of $\boldsymbol{\varrho}$, 
there can be further branches of the moduli space beyond the HB, indicating that 
the HB is singular there. The nature of the effective field theory at the intersection of these branches tells us whether the singularity is resolvable or not. As we observed, the possibility of resolving the singularity is signaled by a residual $U(1)$ gauge group. The associated real FI-term is the resolution parameter.  When such $U(1)$ is absent, the singularity is not a singularity of the total space of the family, and instead signals the presence of a T-brane background, as discussed in  \cite{Collinucci:2016hpz}. 

We have also concluded that our effective theory is  able to detect features of M-theory/type IIA reduction on  threefolds and fourfolds that are obtained by selecting respectively a one-dimensional and two-dimensional subspace of $B_\varrho$.

\

Natural directions for future work include extending our analysis done for families of type A and D to  the E-type families. This will enable the study of universal flops of $\ell=3,4,5,6$ by our methods. Technical issues must be understood in these cases from the QFT side (as the nodes that need to be removed from the E-quivers support non-abelian groups). Another avenue  for exploration (that needs the same QFT issue to be resolved) would be considering stacks of $N$ D2-brane probes, promoting the quiver algebras studied here to their non-abelian counterparts.

Our approach also offers a powerful tool for analyzing simple CY threefold flops, that are realized as subvarieties within the ADE family by selecting a proper curve in the base $B_\varrho$. 
By promoting the coordinate on such a curve to a dynamical field in the 3d theory, one can describe the CY threefold as part of the moduli space of a D2-brane probing that threefold. This must related to the setup of \cite{Cachazo:2001gh} with D3-branes probing such CY threefolds. This point will be explored in more depth in a forthcoming companion paper~\cite{newpaper}.

The techniques developed in this paper could also be applied to Calabi-Yau fourfolds derived from ADE families, for instance by allowing $\boldsymbol{\varrho}$ to depend on two complex variables instead of one. Recently, \cite{Sangiovanni:2024nfz} presented a large class of fourfolds along with their corresponding $\Phi$. The explicit knowledge of  $\Phi$ makes it possible to analyze these spaces using the methods we have outlined in this work.

\section*{Acknowledgments}

First of all, we wish to express our gratitude to Andr\'es Collinucci for the numerous discussions on various aspects of this work. We have benefited from fruitful conversations with Andrea Sangiovanni and Simone Giacomelli.
M.M. and R.V. acknowledge support by INFN Iniziativa Specifica ST\&FI.

\appendix

\section{Relation among $A_1,A_2,A_3,B$ for $\ell=2$ universal flop}\label{App:UnFl2relquiver}
Here we derive the relation \eqref{eq:BA1A2A3equation}, that we used in Section~\ref{Sec:EqFamilyHBUnFl2} for computing the universal flop of $\ell=2$ equation. 

We start by considering the identity:
\be 
A_2A_1A_3= \text{tr}(\mathfrak{M}_0\mathfrak{M}_2\textcolor{red}{\mathfrak{M}_0}\mathfrak{M}_1\mathfrak{M}_0\mathfrak{M}_3)\:.
\ee  
We replace the red-colored $\mathfrak{M}_0$ with: $\textcolor{red}{\mathfrak{M}_0}=\varrho_4\mathbbm{1}-\mathfrak{M}_1-\mathfrak{M}_2-\mathfrak{M}_3$, which comes from \eqref{eq:FtermD4fam3}. 
Since $\text{tr}(\mathfrak{M}_0)=2\varrho_4$ and $\mathfrak{M}_i^2=\varrho_i\mathbbm{1}$, we have:
\begin{flushleft}
$A_2A_1A_3= -\text{tr}(\mathfrak{M}_0\mathfrak{M}_2\mathfrak{M}_3\mathfrak{M}_1\textcolor{red}{\mathfrak{M}_0}\mathfrak{M}_3)-\varrho_1A_2A_3-\varrho_2A_1A_3+\varrho_4\tr(\mathfrak{M}_0\mathfrak{M}_2\mathfrak{M}_1\mathfrak{M}_0\mathfrak{M}_3)$.
\end{flushleft} 
Again, we use \eqref{eq:FtermD4fam3} to replace the colored $\mathfrak{M}_0$. As a result: 
\begin{equation}
\begin{aligned}
A_2A_1A_3= \,&\text{tr}(\mathfrak{M}_0\mathfrak{M}_2\mathfrak{M}_3\textcolor{red}{\mathfrak{M}_1}\mathfrak{M}_2\mathfrak{M}_3)-\varrho_1A_2A_3-\varrho_2A_1A_3+\varrho_1\varrho_3A_2+\\
&+\varrho_3\tr(\mathfrak{M}_0\mathfrak{M}_2\mathfrak{M}_3\mathfrak{M}_1)-\varrho_4\tr(\mathfrak{M}_0\mathfrak{M}_2\mathfrak{M}_3\mathfrak{M}_1\mathfrak{M}_3)+\varrho_4\tr(\mathfrak{M}_0\mathfrak{M}_2\mathfrak{M}_1)A_3\:.
\end{aligned}
\end{equation} 
Finally, we use again\eqref{eq:FtermD4fam3} to replace the $\mathfrak{M}_1$ in red as
$\textcolor{red}{\mathfrak{M}_1}\rightarrow \varrho_4\mathbbm{1}-\mathfrak{M}_0-\mathfrak{M}_2-\mathfrak{M}_3$, obtaining:
\begin{equation}
\begin{aligned}
A_2A_1A_3=\,\,& -\text{tr}(\mathfrak{M}_0\mathfrak{M}_2\mathfrak{M}_3\mathfrak{M}_0\mathfrak{M}_2\mathfrak{M}_3)-\varrho_1A_2A_3-\varrho_2A_1A_3+\varrho_1\varrho_3A_2-\varrho_3\varrho_2(A_3+A_2)\\
&+\varrho_3\textcolor{orange}{\tr(\mathfrak{M}_0\mathfrak{M}_2\mathfrak{M}_3\mathfrak{M}_1)}+\varrho_4\textcolor{blue}{\tr(\mathfrak{M}_0\mathfrak{M}_2\mathfrak{M}_3\mathfrak{M}_2\mathfrak{M}_3)}-\varrho_4\textcolor{purple}{\tr(\mathfrak{M}_0\mathfrak{M}_2\mathfrak{M}_3\mathfrak{M}_1\mathfrak{M}_3)}\\ &+\varrho_4\textcolor{violet}{\tr(\mathfrak{M}_0\mathfrak{M}_2\mathfrak{M}_1)}A_3\:.
\end{aligned}
\end{equation} 
Now, let us compute 
\begin{equation}
\begin{aligned}
\textcolor{violet}{\tr(\mathfrak{M}_0\mathfrak{M}_2\mathfrak{M}_1)}=&\,-\tr(\mathfrak{M}_0\mathfrak{M}_2\mathfrak{M}_3)-\varrho_4A_2-2\varrho_4\varrho_2\\
\textcolor{orange}{\tr(\mathfrak{M}_0\mathfrak{M}_2\mathfrak{M}_3\mathfrak{M}_1)} =&\, \varrho_4\tr(\mathfrak{M}_0\mathfrak{M}_2\mathfrak{M}_1)-\varrho_2A_1-\varrho_1A_2-A_2A_1\\ =&\,-\varrho_4\tr(\mathfrak{M}_0\mathfrak{M}_2\mathfrak{M}_3)-\varrho_4^2A_2-2\varrho_4^2\varrho_2-\varrho_2A_1-\varrho_1A_2-A_2A_1\\
\textcolor{purple}{\tr(\mathfrak{M}_0\mathfrak{M}_2\mathfrak{M}_3\mathfrak{M}_1\mathfrak{M}_3)} =&\, -\textcolor{blue}{\tr(\mathfrak{M}_0\mathfrak{M}_2\mathfrak{M}_3\mathfrak{M}_2\mathfrak{M}_3)}-(A_3+\varrho_3)\tr(\mathfrak{M}_0\mathfrak{M}_2\mathfrak{M}_3)+\varrho_4\varrho_3A_2\\
=&\,-\textcolor{blue}{\left[ \tr(\mathfrak{M}_0\mathfrak{M}_2\mathfrak{M}_3\{\mathfrak{M}_2,\mathfrak{M}_3\})-2\varrho_4\varrho_3\varrho_2
\right]}-(A_3+\varrho_3)\tr(\mathfrak{M}_0\mathfrak{M}_2\mathfrak{M}_3)+\varrho_4\varrho_3A_2 \\
=&\, - \textcolor{blue}{\left[ \tr(\mathfrak{M}_0\mathfrak{M}_2\mathfrak{M}_3)(A_1+\varrho_4^2+\varrho_1-\varrho_3-\varrho_2)-2\varrho_4\varrho_3\varrho_2
\right]}\\ &\, -(A_3+\varrho_3)\tr(\mathfrak{M}_0\mathfrak{M}_2\mathfrak{M}_3)+\varrho_4\varrho_3A_2 
\end{aligned}
\end{equation}
where in the last line, we have applied  $(\mathfrak{M}_2+\mathfrak{M}_3)^2=(\mathbbm{1}\varrho_4-\mathfrak{M}_0-\mathfrak{M}_1)^2$.

Using these relations and the fact that $\sum_i A_i=-2\varrho_4^2$ we can write
\begin{equation}
\begin{aligned}
A_2A_1A_3=\,\,&-\text{tr}(\mathfrak{M}_0\mathfrak{M}_2\mathfrak{M}_3\mathfrak{M}_0\mathfrak{M}_2\mathfrak{M}_3)-(\varrho_1+\varrho_4^2)A_2A_3-(\varrho_2A_3+\varrho_3A_2)(A_1+2\varrho_4^2)\\
&+2\varrho_4\tr(\mathfrak{M}_0\mathfrak{M}_2\mathfrak{M}_3)\left[A_1+\varrho_4^2+\varrho_1-\varrho_2-\varrho_3\right]-4\varrho_4^2\varrho_3\varrho_2
\end{aligned}
\end{equation}  
Finally, using $\text{tr}(\mathfrak{M}_0\mathfrak{M}_2\mathfrak{M}_3\mathfrak{M}_0\mathfrak{M}_2\mathfrak{M}_3)=\left[ \text{tr}(\mathfrak{M}_0\mathfrak{M}_2\mathfrak{M}_3) \right]^2$ and $B\equiv \text{tr}(\mathfrak{M}_0\mathfrak{M}_2\mathfrak{M}_3)$, we obtain \eqref{eq:BA1A2A3equation}.

\section{Universal flop of length $2$ from $SU(2)$ with $4$ flavors}\label{AppD4mirror}

Let us see what happens in the mirror dual of the $D_4$ quiver gauge theory, when a background for $\Phi$ is turned on. The mirror theory is  a $SU(2)$ gauge theory with $4$ flavors; it is realized by a D2-brane probing a stack of 4 D6-branes (plus their images) on top of an orientifold O6-plane. 

Mirror symmetry maps the moment map $\mu$ (above \eqref{eq:P1P2P3P4}) into the meson matrix of the dual theory ($SU(2)$ with $4$ flavors):
\begin{equation*}
\mathcal{M}=\begin{pmatrix} Q^T\tilde{Q}^T & Q^T\epsilon Q\\ -\tilde{Q}\epsilon^{-1} \tilde{Q}^T &-\tilde{Q}Q
\end{pmatrix}
\end{equation*}
\begin{equation}\label{mirrmap}
\updownarrow
\end{equation}\\

Let us consider the Higgs field generating the universal flop of length $\ell=2$, i.e. $\Phi$ given in \eqref{Flop2Phi}.
Turning on such a vev on the worldvolume of the D6-branes will induce the following mass term on the probe D2-brane:
\begin{equation}
\begin{aligned}				
\delta W =\frac{1}{2}\text{Tr}(\Phi \mathcal{M})
 =\,\,&\tilde{Q}_1Q^2+\tilde{Q}_3Q^4-\tilde{Q}_4\epsilon^{-1}\tilde{Q}^T_3+\varrho_1\tilde{Q}_2Q^1\\
 &+\varrho_2(Q^3)^T\epsilon Q^4+\varrho_3\tilde{Q}_4Q^3+\varrho_4(\tilde{Q}_1Q^1+\tilde{Q}_2Q^2)
\end{aligned}
\end{equation}\\
Here we adopt the conventions:
$\epsilon\equiv \epsilon_{ab}\, (\epsilon_{12}=-1),$ $\epsilon^{-1}\equiv \epsilon^{ab}$, $\tilde{Q}\equiv \tilde{Q}^b_i, \quad Q=Q^j_a$.
The deformed superpotential reads:
\begin{equation}\label{superpotdef}
W=\tilde{Q}\Psi Q+\tilde{Q}_1Q^2+\tilde{Q}_3Q^4-\tilde{Q}_4\epsilon^{-1}\tilde{Q}^T_3+\varrho_1\tilde{Q}_2Q^1+\varrho_2(Q^3)^T\epsilon Q^4+\varrho_3\tilde{Q}_4Q^3+\varrho_4(\tilde{Q}_1Q^1+\tilde{Q}_2Q^2)
\end{equation}  
The mass terms in \eqref{superpotdef} deform the CB of the theory. To see this, let us analyze the mass spectrum of matter fields on a generic point of the CB. Here the gauge group is broken to its Cartan and 
$\Psi=\psi_3\sigma_3$ with $\sigma_3$ denoting the third Pauli matrix.
The mass matrix is then given by 
{\footnotesize \begin{equation*}
m=\begin{pmatrix} \psi_3\sigma_3+\varrho_4\mathbbm{1}_2&\mathbbm{1}_2&0&0&0&0&0&0\\ \varrho_1\mathbbm{1}_2&\psi_3\sigma_3+\varrho_4\mathbbm{1}_2&0&0&0&0&0&0\\ 0&0&\psi_3\sigma_3&\mathbbm{1}_2&0&0&0&\epsilon^{-1}\\0&0&\varrho_3\mathbbm{1}_2&\psi_3\sigma_3&0&0&-\epsilon^{-1}&0\\0&0&0&0&-\psi_3\sigma_3-\varrho_4\mathbbm{1}_2&-\varrho_1\mathbbm{1}_2&0&0\\0&0&0&0&-\mathbbm{1}_2&-\psi_3\sigma_3-\varrho_4\mathbbm{1}_2&0&0\\0&0&0&-\varrho_2\epsilon&0&0&-\psi_3\sigma_3&-\varrho_3\mathbbm{1}_2\\0&0&\varrho_2\epsilon&0&0&0&-\mathbbm{1}_2&-\psi_3\sigma_3\end{pmatrix}
\end{equation*}}
and the mass term reads $W_m\equiv  \frac{1}{2}(\tilde{Q}, Q^T)\cdot m\cdot (Q^T, \tilde{Q})^T$.

From it, we can deduce the monopole equation for the CB after turning on $\Phi$ \cite{Bullimore:2015lsa}:
\begin{equation}\label{cbeq}
V_+V_-= \frac{\sqrt{\text{det}(m)}}{\psi_3^4}=\frac{((-\psi_3^2+\varrho_1-\varrho_4^2)^2-4\varrho_4^2\psi_3^2)((-\psi_3^2+\varrho_3-\varrho_2)^2-4\varrho_2\psi_3^2)}{\psi_3^4}
\end{equation}
At the denominator, we find the contribution from the W bosons' masses, which are left unchanged by the deformation. 
Out of $V_{\pm}, \psi_3$ we construct the Weyl invariant operators:
\begin{equation}
z\equiv -\psi_3^2, \quad y\equiv \frac{V_++V_-}{2}+\frac{(\varrho_2-\varrho_3)(\varrho_1-\varrho_4^2)}{\psi_3^2}, \quad x\equiv \psi_3\frac{V_+-V_-}{2}.
\end{equation}
Under such identifications, we have: 
\begin{equation}
x^2+z\left(y+\frac{(\varrho_2-\varrho_3)(\varrho_1-\varrho_4^2)}{z}\right)^2=-\psi_3^2V_+V_-\:.
\end{equation}
The equation \eqref{cbeq} then implies:
\begin{equation}
\begin{aligned}
x^2+zy^2=&\frac{((z+\varrho_1-\varrho_4^2)^2+4\varrho_4^2z)((z+\varrho_3-\varrho_2)^2+4\varrho_2z)-(\varrho_2-\varrho_3)^2(\varrho_1-\varrho_4^2)^2}{z}\\ &-2y(\varrho_2-\varrho_3)(\varrho_1-\varrho_4^2)\:.
\end{aligned}
\end{equation}
This is the expected equation \eqref{eq:Flop2Eq} for the universal flop of length $\ell=2$.

The computation of the universal flop equation on this side of the  duality, gives the same result as computed with the mirror dual \eqref{eq:UnivFlFromQuiverXYZ} (up to the coordinate change \eqref{eq:Fl2ChangeCoord}). This is expected as we believe in mirror symmetry. Of course, the match is perfect if the duality map is settled appropriately, and this happens when the moment map $M$ on this side is mapped by mirror symmetry to $\mu$ in the equation above \eqref{eq:P1P2P3P4} (moreover, we are also assuming that the abelian mirror maps as in Section~\ref{Sec:A1SingAnd3dMirro} are correct). One can also prove that, up to a proper rescaling, the quantum monopole relations satisfied by the elements of $\mu$ reproduce the right relations the meson matrix $M$ should fulfill.

\providecommand{\href}[2]{#2}

\end{document}